\documentclass[%
aps,
prx,
reprint,
superscriptaddress,
nofootinbib,
amsmath,
amssymb
]{revtex4-2}
\usepackage{
    phfqit,
    graphicx,
    multirow,
    tabularx,
    placeins, 
    nicefrac, 
    hyperref, 
    threeparttable, 
    enumitem,
    makecell,
    comment,
}

\usepackage[print-unity-mantissa=false]{siunitx}
\usepackage{mathtools}\mathtoolsset{centercolon}
\usepackage{fixmath}
\usepackage{phfparen}
\usepackage{mathcommand}
\usepackage{stmaryrd}
\LoopCommands{\lettersAll{CG}{DCS}{MSPE}{NLL}{SE}}{\declaremathcommand#2{\textnormal{#1}}}
\LoopCommands{
{pr}{wt}
}{\DeclareMathOperator#2{#1}}

\usepackage{algorithm}
\usepackage{algpseudocodex}
\usepackage{amsthm}
\usepackage[capitalize]{cleveref}
\crefname{section}{Section}{Sections}
\Crefname{section}{Section}{Sections}

\theoremstyle{definition}
\newtheorem{defn}{Definition}

\theoremstyle{remark}
\newtheorem{rem}{Remark}

\crefname{rem}{Remark}{Remarks}
\Crefname{rem}{Remark}{Remarks}

\usepackage[dvipsnames]{xcolor}
\hypersetup{
    colorlinks,
    linkcolor={blue!50!black},
    citecolor={blue!50!black},
    urlcolor={blue!80!black}
}

\usepackage[caption=false]{subfig} 
\DeclarePairedDelimiterX\pbraket[2]{\langle\!\langle}{\rangle\!\rangle}{#1 \delimsize\vert #2}

\begin{document}
\setcounter{secnumdepth}{0}

\title{Error mitigation for logical circuits using decoder confidence }

\newcommand{\qmaddress}{\affiliation{Quantum Motion, 9 Sterling Way, London N7 9HJ, United Kingdom}}
\newcommand{\oxaddress}{\affiliation{Department of Materials, University of Oxford, Parks Road, Oxford OX1 3PH, United Kingdom}}
\newcommand{\osaddress}{\affiliation{Center for Quantum Information and Quantum Biology, The University of Osaka, 1-2 Machikaneyama, Toyonaka 560-0043, Japan}}

\author{Maria Dincă}
\email{maria.dinca@materials.ox.ac.uk}
\oxaddress
\author{Tim Chan}
\email{timothy.chan@materials.ox.ac.uk}
\oxaddress
\osaddress
\author{Simon C. Benjamin}
\oxaddress
\qmaddress

\date{\today}

\begin{abstract}
\textbf{ABSTRACT}\par\medskip
Fault-tolerant quantum computers use decoders to monitor for errors and find a plausible correction. A decoder may provide a \emph{decoder confidence score} (DCS) to gauge its success. We adopt a \emph{swim distance} DCS, computed from the shortest path between syndrome clusters. By contracting tensor networks,
we compare its performance under phenomenological noise
to the well-known \emph{complementary gap} and find that
both reliably estimate the logical error probability (LEP) in a decoding window. We explore ways to use this to mitigate the LEP in entire logical circuits. For shallow circuits, we just abort if any decoding window produces an exceptionally low DCS: for one window of a distance-13 surface code under circuit-level noise, rejecting a mere $0.1\%$ of possible DCS values improves the window's mean LEP by more than 5 orders of magnitude. For larger algorithms comprising up to billions of windows, DCS-based rejection remains effective for enhancing observable estimation. Moreover, one can use the DCS to assign each circuit's output a unique LEP, and use it as a basis for maximum likelihood estimation. This can reduce the effects of noise by an order of magnitude at no quantum cost; we explore the combination of these methods.

\end{abstract}

\maketitle

\section{Introduction} \label{part_0}

In fault-tolerant quantum computing, we perform a series of measurements that can reveal information about the occurrence of errors on the system of qubits, while preserving the logical state. The set of observations generated by these measurements forms a syndrome. A key task in quantum error correction (QEC)~\cite{shor_scheme_1995, fowler_surface_2012} is to map the syndromes observed on the error correcting code to the suspected physical error mechanisms which might have produced them, and consequently suggest plausible corrections. This process is known as decoding.

There has been recent interest in extracting more information from the decoder than just a correction;
namely, an estimate of how likely it is for the most plausible correction to be actually correct.
In the past,
this estimate has been called `soft output'~\cite{Chen2026,meister_efficient_2024,Lee2025c}
or `soft information'~\cite{bombin_fault-tolerant_2024,deMartiiOlius2024,Gidney2025,Pattison2025,Akahoshi2025,Sunami2025},
but we opt for the name `decoder confidence score' (DCS) as it is more descriptive.
It has been identified that additional information of this kind can be used in
code concatenation~\cite{Poulin2006,Goto2013,Gidney2025,meister_efficient_2024},
defect identification in silicon-spin quantum processors~\cite{siegel_snakes_2025},
magic state preparation~\cite{Smith2024,SalesRodriguez2025,gidney_magic_2024,Chen2026,Sahay2025},
circuit sampling~\cite{meister_efficient_2024, Smith2024,bluvsteinLogicalQuantumProcessor2024},
lattice surgery runtime reduction~\cite{Akahoshi2025},
hierarchical decoding~\cite{Shutty2024,Toshio2025},
distributed quantum computing~\cite{Sunami2025},
fusion-based computing~\cite{Birchall2026,Staples2026},
and decoder runtime reduction~\cite{Oberoi2026,Mishima2026}.

A significant opportunity provided by the DCS lies in quantum error mitigation (QEM). QEM was developed during the NISQ regime as a family of methods that reduce noise-induced bias by combining multiple noisy circuit executions with classical post-processing, typically trading qubit overhead for increased sampling cost \cite{cai_quantum_2023}. Representative approaches include zero-noise extrapolation (ZNE) \cite{liEfficientVariationalQuantum2017,temmeErrorMitigationShortdepth2017, giurgica-tironDigitalZeroNoise2020}, probabilistic error cancellation (PEC) \cite{temmeErrorMitigationShortdepth2017,endo_practical_2018,mariExtendingQuantumProbabilistic2021,moske_random-matrix_2026}, symmetry verification \cite{bonet-monroigLowcostErrorMitigation2018,mcardleErrormitigatedDigitalQuantum2019,caiQuantumErrorMitigation2021}, purification-based methods \cite{hugginsVirtualDistillationQuantum2021, koczorExponentialErrorSuppression2021,obrienPurificationbasedQuantumError2023}, and learning-based estimators \cite{czarnikErrorMitigationClifford2021,strikisLearningbasedQuantumError2021}, with recent experimental confirmations of their scalability \cite{vandenbergProbabilisticErrorCancellation2023,kim_evidence_2023, leviatan_resolving_2026}.

Several works have argued that these ideas remain relevant and should be adapted to the early fault-tolerant regime, where limited code distances lead to non-negligible logical error rates \cite{suzuki_quantum_2022, aharonov_importance_2025}. One of the most straightforward extensions is ZNE via Pauli-error injection on physical qubits, which was demonstrated experimentally on the repetition and surface codes \cite{zhang_demonstrating_2025}. Ref. \cite{jeon_quantum_2026} later provided a theoretical justification, proving more generally that any linear QEM method can be applied at the physical layer of an error correcting code without modifying the subsequent decoder. Ref. \cite{dutkiewicz_error_2024} developed a QEM strategy suitable for early fault tolerance beyond expectation value estimation problems, based on PEC-like usage of logical noise characterisation; its efficiency relies on low logical gate error rates. While all these techniques merge smoothly with QEC, a complementary approach is to exploit the additional resources offered by QEC, such as using the code distance as a noise scaling parameter in ZNE \cite{wahl_zero_2023,umbrarescu_infinite_2026}.
We address how to include DCS information in QEM protocols within the latter line of thought.

Typically, fault-tolerant quantum computing involves decoding in a series of windows:
batches of syndromes collected in a finite amount of time
across a small region of space are decoded at once.
A typical window extent is a small multiple of the code distance~\cite{Dennis2002}.
The DCS is particularly useful here,
since each window can be associated with its own DCS value,
thus providing a localised failure probability for each component of the quantum computation.

Two existing DCSs that have been studied for the surface code are
the complementary gap and a cheaper alternative which we call the swim distance.
The complementary gap was first introduced as `distinguishability' in~\cite{Hutter2014},
before being studied further in Refs.~\cite{bombin_fault-tolerant_2024,Gidney2025}.
The swim distance was introduced simultaneously in Refs.~\cite{Smith2024,meister_efficient_2024}
as a faster alternative to the complementary gap
(it has a lower time complexity);
however,
neither reference provides an accuracy comparison against the complementary gap. We note that the term `cluster gap' has also been used for this latter DCS~\cite{Toshio2025}.

In this paper, we refine the picture of the DCS with three observations.
First, we use tensor network methods
to compare the accuracy of the complementary gap and the swim distance
under phenomenological noise.
While we find the complementary gap to be somewhat more accurate,
both DCSs are accurate proxies for the success probability.
For the rest of the paper,
we employ the swim distance as it is faster to compute,
and assume a circuit-level noise model.
Second, we identify characteristics from the statistical distribution of the DCS
that make it a useful resource for large-scale applications.
Third, we explore this utility for quantum error mitigation in logical circuits.
In the case of a shallow circuit -- such as logical state preparation --
simply aborting when a poor DCS occurs is very effective.
We then consider the case of an entire algorithm,
which may involve a vast number of decoding windows.
The strategy of aborting unpromising circuits remains tenable,
but we find that another option is attractive:
estimate the whole-circuit logical error probability using its DCS record,
then perform maximum likelihood estimation using the estimated risk. Depending on target performance,
a combination of methods may be optimal. This is shown in \cref{fig:MLE_abort}b,d as a primary result of this work.

The ``Results" section first introduces the prerequisite concepts and compares the complementary gap and the swim distance, then shows that, even for deep circuits, the DCS provides rich information, which can be used in logical-level QEM, as exemplified by the Hubbard-model numerical resource estimation. Related work, limitations, and future directions are elaborated on in ``Discussion", and the computational and simulation procedures are described in ``Methods".

\section{Results}\label{sec:background}
\subsection{The Role of a Decoder}\label{sec:the_role_of_a_decoder}
In this work, we focus on the surface code,
for which the $X$ and $Z$ errors can be dealt with separately on independent decoding graphs.
We consider decoding only one type (either $X$ or $Z$) of error;
the results for the other type of error are analogous.
In error correction,
nondestructive parity measurements are performed on physical qubits
and the combination of their results is called the syndrome.
Error chains (configurations of errors on the physical qubits)
are detected only through their syndrome,
but different error chains can give rise to the same syndrome.
Additionally,
each error chain belongs to one of two logical equivalence classes (LECs)
that uniquely specifies its effect on the logical information.
A decoder deduces from the syndrome
a plausible LEC for the error chain that caused the syndrome.

We will mainly be concerned with decoders that deduce an explicit plausible error chain.
What matters is
not whether the deduced error chain equals the true error chain,
but whether they belong to the same LEC.
If they are in different LECs,
then the decoding procedure results in a logical error.
Therefore, a decoder effectively makes guesses --
and while even the ideal decoder makes the best possible guess,
it may nevertheless lead to a logical error.
When choosing a decoder,
there is a trade-off between the accuracy with which the likeliest LEC is found and the time it takes to decode.

\begin{figure}[tbp]
    \centering
    \begin{tabular}{cc}
        \includegraphics[width=\linewidth]{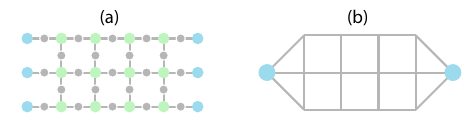}
    \end{tabular}
    \caption{\textbf{Surface-code decoding graphs before and after merging equivalent boundary nodes.} (a) The $X$ decoding graph for the $\llbracket n,k,d_X,d_Z\rrbracket=\llbracket23,1,5,3\rrbracket$ unrotated surface code under code-capacity noise. Detectors are green circles, boundary nodes are blue circles, and data qubits are emphasised by grey circles on the corresponding edges. In this noise model, but not necessarily in others, each detector corresponds to a $Z$ stabiliser and each edge corresponds to a data qubit. (b) For later convenience, equivalent boundary nodes are merged, so that there is only one blue boundary node for each unique boundary of the surface code.
    Here, we no longer show detectors nor data qubit circles.}\label{fig:to_decoding_graph}
\end{figure}

More formally,
the decoding problem involves the decoding graph $G =(V_\d \cup V_\b, E)$,
with the set $V_\d$ of detectors and the set $V_\b$ of boundary nodes such that $V_\d \cap V_\b =\varnothing$.
Each boundary node in $V_\b$ represents a unique boundary of the surface code
(so for a memory experiment, $|V_\b| =2$).
An example of how $G$ relates to its code is shown in \cref{fig:to_decoding_graph}.
Each edge in $E$ represents an error mechanism that flips the one or two detectors it is incident to,
and has a weight $\log_{10} \frac{1 -p}{p}$ that decreases
as the probability $p$ of that error mechanism increases.
Define the weight $\wt(F)$ of any subset $F \subseteq E$ as the sum of the weights of all edges in that subset. 
The error $\mathbb E \subseteq E$ represents those mechanisms that actually occur,
from which we define the syndrome $\mathbb S \subseteq V_\d$ as
the detectors that $\mathbb E$ flips an odd number of times.
Given $G$ and $\mathbb S$,
the decoder outputs a correction $\mathbb C =\textsc{Decode}(G, \mathbb S) \subseteq E$
that will have the same syndrome as $\mathbb E$.
We say the decoder is {\em successful} if $\mathbb C$ is logically equivalent to $\mathbb E$.
\begin{defn}\label{defn:log_success_odds}
    The \emph{logical error probability} (LEP), denoted $p_\L$,
    is the probability that the decoder is not successful,
    conditioned on $\mathbb S$.
    The \emph{log success odds} (or \emph{log-likelihood ratio of success}) is given by
    \begin{equation}\label{eq:log_success_odds_definition}
        \lambda :=\log_{10} \frac{1 -p_\L}{p_\L}.
    \end{equation}
\end{defn}
\noindent
The log success odds decreases as $p_\L$ increases.
In addition to the LEP,
we define the following.
\begin{defn}\label{defn:ler}
    When we have performed multiple runs of the same logical circuit,
    the \emph{logical error rate} (LER), denoted $R$,
    is the portion of executions with a logical error.
\end{defn}
\noindent The LER converges to the mean LEP
as the run count grows.

\subsection{Decoder Confidence Scores}\label{sec:decoder_confidence_scores}
In addition to the correction,
the decoder can also output the following quantity
that scores how likely it believes success to be.
\begin{defn}\label{defn:dcs}
    Given $G$ and $\mathbb S$,
    the \emph{DCS} is an estimator
    of some monotonic function $\alpha(\lambda)$ of $\lambda$.
    We denote its value as $\phi$.
\end{defn}
\noindent
The quantity $p_\L$ (or $\lambda$) can be exactly computed
by contracting a tensor network~\cite{piveteau_tensor-network_2024};
this was analysed recently in Ref.~\cite{Chen2025a}.
However,
this method cannot be used as a real-time DCS as its time complexity is exponential.
The following two definitions specify DCSs that run in polynomial time.

\begin{figure}[tbp]
    \centering
    \begin{tabular}{cc}
        \includegraphics[width=\linewidth]{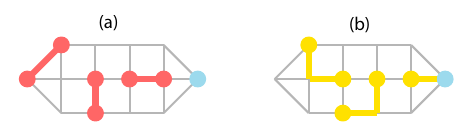}
    \end{tabular}
    \caption{\textbf{Illustration of complementary gap
    on the decoding graph $G$ from \cref{fig:to_decoding_graph}b.}
    The modified graph $G'$, shown here in grey,
    has only one remaining boundary node, shown as a blue dot. (a) Initial decoding. (b) Decoding in the opposite logical equivalence class.
    The syndrome $\mathbb S$ ($\mathbb S'$) is the set of red (yellow) dots.
    The correction $\mathbb C$ ($\mathbb C'$) is the set of red (yellow) edges.
    The complementary gap is the weight difference between these two corrections;
    here it is $5 -3 =2$ if all edges have weight 1.}
    \label{fig:gap_illustration}
\end{figure}

\begin{defn}\label{defn:complementary_gap}
The \emph{complementary gap} is the difference in weight between
the original correction and the complementary correction
(the correction output by the decoder
when forced to propose within the opposite LEC).
\end{defn}
\noindent
We can force the decoder to propose the opposite LEC
by forcing the left boundary node to flip
a number of times whose parity is opposite to that of the original correction.
This is described in \cref{alg:complementary_gap}
and illustrated by example in \cref{fig:gap_illustration}.

\begin{algorithm}[H]
\caption{Pseudocode to calculate the complementary gap of a memory experiment.}
\label{alg:complementary_gap}
\begin{algorithmic}
    \State $\mathbb C \gets \Call{Decode}{G, \mathbb S}$
    \State $G' \gets G$ but where the left boundary node $v_\L$ is a detector
    \State $\mathbb S' \gets$ the syndrome of $\mathbb C$ on $G'$
    \State flip whether or not $v_\L$ is included in $\mathbb S'$
    \State $\mathbb C' \gets$ \Call{Decode}{$G', \mathbb S'$}
    \State \Return $\wt(\mathbb C') -\wt(\mathbb C)$
\end{algorithmic}
\end{algorithm}

\begin{figure}[tbp]
    \centering
    \includegraphics[width=0.6\linewidth]{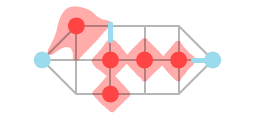}
    \caption{\textbf{Illustration of swim distance on the decoding graph $G$ from \cref{fig:to_decoding_graph}b.}
    Here, the swim distance is the sum of the lengths of the blue lines:
    $2 \times 0.5 =1$ if all edges have weight 1.
    Clusters are shaded in red.}\label{fig:swim_distance_illustration}
\end{figure}

\begin{defn}\label{defn:swim_distance}
    For any clustering decoder,
    the \emph{swim distance}
    is the shortest path length in the decoding graph from one boundary to the other
    after zeroing the weights of all edges inside clusters.
\end{defn}
\noindent
Note a cluster can contain a fraction of an edge,
in which case only that fraction of its weight is set to zero.
\Cref{fig:swim_distance_illustration} illustrates the swim distance
for the same example as in \cref{fig:gap_illustration}
and shows that its value can differ from the complementary gap.
The name `swim distance' is based on the following analogy:
the decoding graph is a river,
the two boundaries are riverbanks,
and each cluster is an island in the river;
then,
the swim distance is the shortest distance one must swim to cross the river.

The complementary gap and the swim distance
are not the only DCSs that exist.
For instance,
Ref.~\cite{bluvsteinLogicalQuantumProcessor2024} considers simply the weight of the original correction,
Ref.~\cite{Lee2025c} defines two families of scores based on how much of the decoding graph is covered by clusters,
and Refs.~\cite{bluvsteinLogicalQuantumProcessor2024,English2025} define a score based on $|\mathbb{S}|$ alone.
Nevertheless,
the complementary gap remains considerably more effective
at reducing the LER in abort protocols~\cite{Lee2025c}.
We choose to analyse the complementary gap for this reason,
and the swim distance because it has a very strong theoretical motivation~\cite{meister_efficient_2024}.
For the numerics in this paper,
we decode using the PyMatching implementation~\cite{Higgott2025} of Minimum Weight Perfect Matching (MWPM),
calculate the complementary gap with in-house code,
and calculate the swim distance using the implementation provided by Ref.~\cite{Chen2026}.

\subsection{Decoder Confidence Scores are an Excellent Indicator of Logical Error Risk}\label{part_1}
\begin{figure}[tbp]
    \centering
    \includegraphics[width=0.95\linewidth]{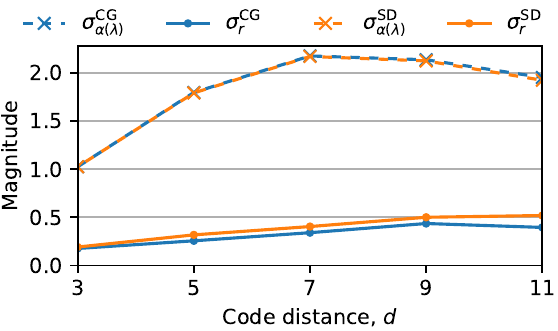}
    \caption{\textbf{Comparison of decoder confidence score (DCS) variation sources across code distance.} Blue data correspond to the complementary gap and orange data to the swim distance.
    The first source $\sigma_{\alpha(\lambda)}$
    is due to the variation of the log success odds,
    and is almost the same between the two DCSs.
    The second source $\sigma_r$
    comes from the approximate nature of each DCS.
    Results are obtained using the Minimum Weight Perfect Matching (MWPM) decoder
    on the unrotated surface code under a perturbed phenomenological noise model
    (described under ``Perturbed Phenomenological Noise Model" in the ``Methods" section).}\label{fig:source_of_dcs_variation_perturb0.3}
\end{figure}

In this section we verify that both the complementary gap and the swim distance
are accurate enough to be useful in error mitigation;
this is formalised as follows.
The log success odds $\lambda$ can be modelled as a random variable
whose variation throughout a typical quantum calculation
is characterised by its standard deviation $\sigma_\lambda$.
Since the DCS $\phi$ is an inexact estimator of $\alpha(\lambda)$,
it can be modelled as
\begin{equation}\label{eq:dcs_as_sum_of_2_rvs}
    \phi =\alpha(\lambda) +r,
\end{equation}
where $r$ is another random variable whose standard deviation $\sigma_r$ represents the DCS inaccuracy.
For $\phi$ to be useful,
it must be able to resolve variations in $\lambda$ between different decoding instances,
i.e., $\sigma_{\alpha(\lambda)} =\alpha'(\lambda) \sigma_\lambda$ must be large compared to $\sigma_r$.
\Cref{fig:source_of_dcs_variation_perturb0.3}
shows this is indeed the case,
for both the complementary gap and the swim distance.
Across all code distances considered,
the complementary gap has a somewhat smaller value of $\sigma_r / \sigma_{\alpha(\lambda)}$ than the swim distance,
which suggests it is a more accurate DCS.
``Calculating the Sources of Variation of the DCS"
describes how we obtained these quantities from our simulations.

While $\sigma_r$ is smaller than $\sigma_{\alpha(\lambda)}$,
it is not much smaller.
The question of imperfect calibration arises:
what is the effect of the DCS inaccuracy on
the (early) fault-tolerant applications presented in the next sections?
Answering this would require calculating the log success odds
for higher code distances under a more realistic noise model than that used in \cref{fig:source_of_dcs_variation_perturb0.3},
but we were unable to do so due to numerical imprecision during tensor network contraction.
In the numerical results below,
we provide a speculative analysis of the DCS uncertainty under a range of assumptions.

We next explore the statistical properties of the DCS distribution which make it a valuable resource
for error mitigation at both
the single-window level, in ``Single-Decoding-Window Statistics",
and the full-application level, in ``Many-Decoding-Window Statistics".
Except for the sliding-window study in ``Discussion", throughout the rest of this paper, 
we focus on the swim distance DCS,
and we will henceforth generally refer to it as `the DCS' and to its value as $\phi$.
We will refer to single-window LEPs and logical errors by lowercase symbols ($p_\L, x$)
and to their multi-window counterparts by uppercase symbols ($P_\L, X$).
We emulate circuits using Stim~\cite{Gidney2021a}
and assume circuit-level noise with a physical error rate $p_\text{phys}=10^{-3}$.
Our approach to estimating DCS performance in the context of full logical circuits is based on allocating a suitable number of decoding windows.
We do not adjust according to e.g.\ the specific decoding needs of logical gates, leaving that for future studies.

\subsection{Single-Decoding-Window Statistics}
\label{sec:single_decoding_window_statistics}
\begin{figure*}[tbp]
    \centering
    \includegraphics[width=\textwidth]{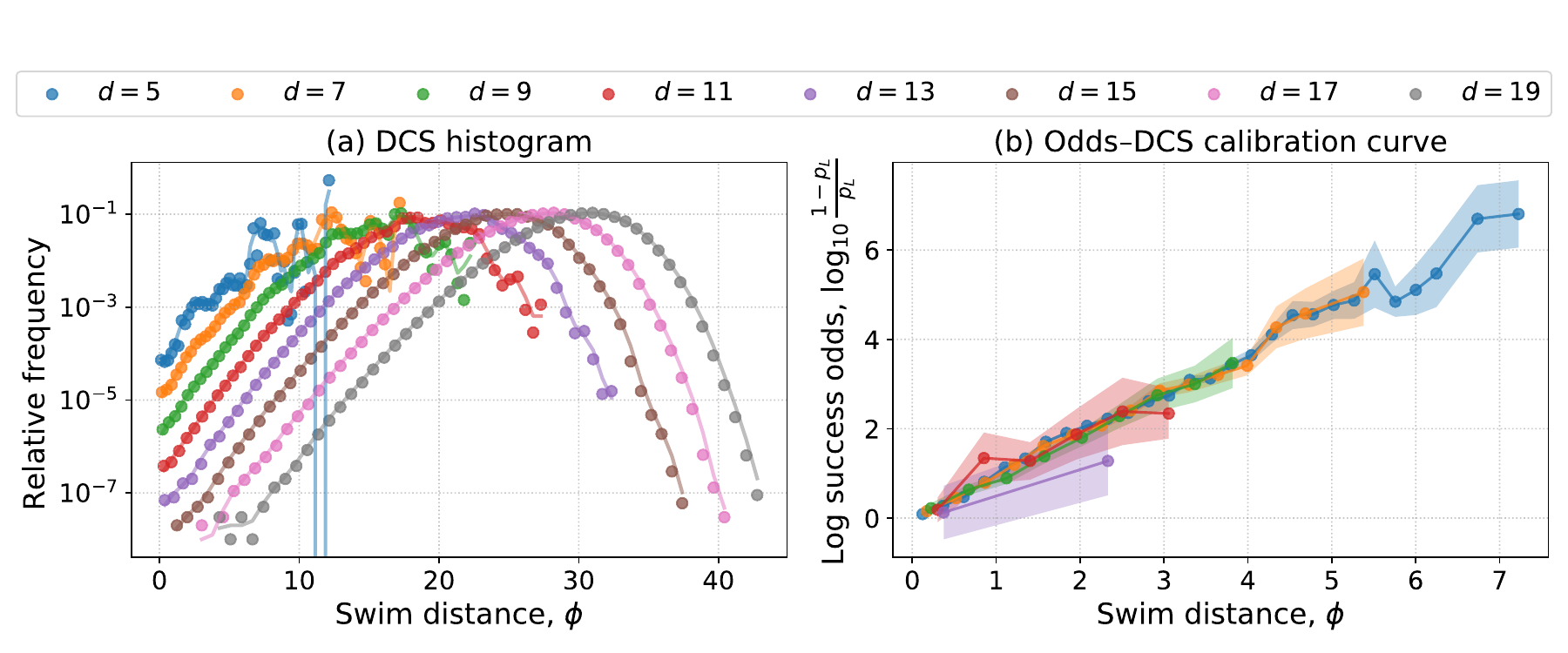}
    \caption{\textbf{Single-decoding-window statistics.}
    (a) Histograms of the decoder confidence score (DCS; here, swim distance) value $\phi$,
    normalised by number of shots. Under circuit-level noise, $10^8$ memory experiments were run for all code distances.
    The $\phi$ values are grouped in 50 bins of equal linear width.
    (b) The log success odds conditioned on $\phi$, plotted against $\phi$.
    The upper and lower bounds on the Bernoulli variable $p_\L$, representing the window logical error probability, are given by the Wilson score interval.
    Only datapoints with $p_\L >0$ are shown.}\label{fig:calibration_curve}
\end{figure*}

We emulated memory experiments in the rotated surface code
for odd code distances $d \in \{5, \dots, 19\}$;
these assess how well one logical qubit stays free from logical error
after $d$ stabiliser measurement cycles.
We decode each experiment in a single window and calculate $\phi$ for that window.
The distribution of $\phi$ is shown in \cref{fig:calibration_curve}a.
Each bin in this histogram contains a number of failed experiments (where a logical error occurred)
and successful (i.e.\ error-free) experiments,
which we use to calculate an empirical log success odds from \cref{eq:log_success_odds_definition}.
\Cref{fig:calibration_curve}b plots this empirical log success odds against $\phi$.
By fitting a straight line to this data (see the Methods subsection ``DCS-Log Success Odds Calibration" for discussion),
we obtain a calibration curve that assigns a LEP $p_\L$ to each $\phi$ value.
These curves can be used in future experiments for the same code distance and noise model.
 
Despite having distinctive peaks, the single-window distributions in \cref{fig:calibration_curve}a provide a wide variety of LEPs if enough shots are taken.
Even in the very early fault-tolerant era, the implementation of a quantum algorithm that is both useful and feasible involves a few hundred logical qubits and thousands of logical operation layers, which translates to billions of decoding windows.
As technology matures further, this number will rise steeply.
In the next subsection, we consider the total LEP of an experiment comprising many windows.

\subsection{Many-Decoding-Window Statistics}
\label{sec:many_decoding_window_statistics}
To simplify analysis,
we assume $(p_\L, x)$ for each window is independent;
this simplification is discussed in ``Discussion".
In this case,
the total LEP $P_\L$ for an experiment comprising $N$ decoding windows,
each with LEP $p_{\L,i}$,
is the probability that an odd number of them were wrongly decoded:
\begin{equation}\label{ler_many_windows}
    P_\L = \frac12 `\bigg[ 1 - \prod_{i=1}^{N} (1-2p_{\L,i}) ].
\end{equation}
A valid concern regarding the usefulness of DCSs for error mitigation is scalability:
does the distribution of the random variable $P_\L$ remain usefully rich in information as the size of the computation increases?
One might imagine that, for large enough computations, $P_\L$ converges to a sharp value, so that
every run with a given code distance and (very large) number of decoding windows $N$ will be found to have a total LEP $P_\L$ of,
for example,
almost exactly $5\%$.
In that case, the useful information gained by evaluating $P_\L$ for a given circuit is negligible -- it is a foregone conclusion.

In fact,
that is not the case.
In Supplementary Note 2
we analyse the distribution of $P_\L$
and find that it remains broad (and its variance grows) as $N$ increases,
if the code distance is adjusted to target a fixed LER.
However, it could still be the case that, while the variance grows with $N$,
the effective range of LEPs spans fewer orders of magnitude. 

\begin{figure*}[tbp]
    \centering
    \includegraphics[width=\textwidth]{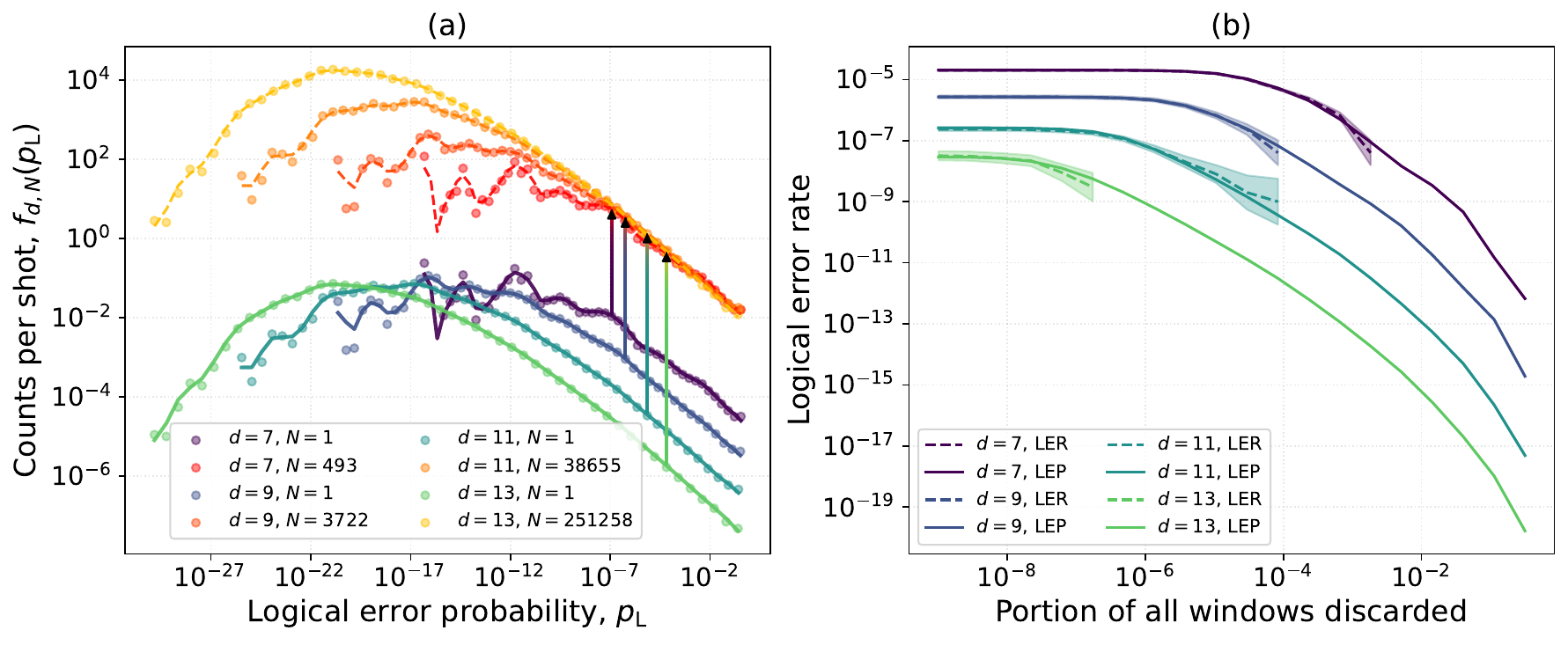}
    \caption{\textbf{Equivalence of different code distances.} (a) Histogram of the calibrated logical error probabilities (LEPs), normalised by the number of shots, for various numbers of decoding windows in the circuit. The LEPs are grouped in 50 bins of equal logarithmic width. At fixed code distance $d$, for circuits comprising $N$ windows, we define $f_{d,N}(p_\L)$ as the average number of constituent windows whose LEP lies in the bin centred on $p_\L$. The case $N=1$ (viridis colourmap) corresponds to the normalised LEP histogram for single-window memory experiments. When more windows are added (magma colourmap), all bins gain population at the same rate: $f_{d,N}(p_\L)=Nf_{d,1}(p_\L)$. It is possible to adjust $N$ such that circuits defined by different $(N,d)$ pairs have the same mean $P_\L$; in this case, the seemingly linear segments of the scaled histograms overlap,
    as indicated by the upward arrows. (b) The calculated mean LEP (solid) and observed logical error rate (LER; dashed) in a cohort of single-window decoding instances when a fraction of highest-$p_\L$ shots are discarded. The curves for $d \in `{11, 13}$ were obtained with $10^9$ shots.}\label{fig:risk_distribution}
\end{figure*}

To investigate this,
we generalise \cref{fig:calibration_curve}a to circuits of $N \ge 1$ decoding windows and plot the result in \cref{fig:risk_distribution}a,
where the $x$-axis has additionally been converted to a LEP via the calibration curve.
Each single-window LEP histogram is scaled by a factor $N$
such that the mean LEP of an $N$-window circuit equals \num{1e-2}. The evident overlap of the scaled histograms for $p_\L > 10^{-7}$ means that the same number of high-$p_\L$ windows will be detected for different $(N,d)$ pairs. These rare, high-risk events are precisely the ones which make the dominant contribution to the overall $N$-window circuit LEP $P_\L$.
\begin{rem}\label{rem:rescale_distributions}
    Because the non-overlapping parts of the scaled histograms do not contribute much to the full-circuit LEP,
    the $P_\L$ distribution for a numerically intractably large pair $(N,d)$ can be approximately inferred from the $P_\L$ distribution for a smaller, testable pair $(N',d')$ yielding the same mean LEP.
\end{rem}
The fact that the low-$p_\L$ windows bring little contribution to $P_\L$ can be seen from \cref{fig:risk_distribution}b, where we consider the scenario of discarding a fraction of the single-window shots with the highest $p_\L$ values. The figure shows both the calculated LEP and the empirically observed LER; these lines agree well.
For $d=13,$
it is possible to improve the LER by one (five) orders of magnitude by discarding a fraction \num{1e-7} (\num{1e-3}) of the total shots.
Finally, \cref{ler_many_windows} gives us an exact expression for one realization of an $N$-window LEP; in the low error rate and sufficient sampling limit, it approximates to $P_\L \approx \sum_{i=1}^N p_{\L,i} \approx N \bar{p}_{\L}$, where $\bar{p}_{\L}$ is the single-window mean LEP. Therefore, since \cref{fig:risk_distribution}b shows that the low-$p_\L$ windows bring little contribution to their single-window mean $\bar{p}_{\L}$, we deduce that they contribute little to most $N$-window LEPs $P_\L$.

We next look at potential applications of
the DCS in QEM for logical circuits in the early fault-tolerant era.
In ``Abort Protocol", we present an abort protocol which halts the execution of a deep quantum circuit based on the \emph{instantaneous} DCS.
``DCS-Based Maximum Likelihood Estimation" explains an MLE procedure based on the \emph{whole-circuit} LEP. In ``Basic DCS-Agnostic Logical QEM", we introduce a baseline protocol for logical-level QEM in the absence of any DCS information.
In the numerical results below,
we combine both methods,
applying MLE on increasingly filtered results
of an expectation value estimation task. The substantial enhancements which we observe represent one of the key aspects of this work.
We conclude with an approximate analytical model
for a distance-19 surface code,
finding that its behaviour closely matches our earlier empirical analyses.

\subsection{Abort Protocol}\label{sec:shallow}
Suppose we choose the following protocol for executing a logical circuit:
abort execution if the DCS associated with \emph{any} decoding window is below a pre-defined threshold.
\Cref{fig:risk_distribution}b immediately leads us to the performance of this approach.
Let $\rho$ be the fraction of windows below the DCS threshold,
and $p_\text{kept}$ be the mean LEP of the retained windows.
These quantities correspond respectively to the $x$ and $y$ axes of \cref{fig:risk_distribution}b.
If an $N$-window circuit successfully completes,
then the expected whole-circuit LEP is approximately $P_\text{kept} = \frac12[1-(1-2p_\text{kept})^N] \approx Np_\text{kept}$.
As noted, $p_\text{kept}$ can be many orders of magnitude lower than when everything is retained; therefore, so can $P_\text{kept}$.
The caveat is that the probability of successfully completing falls as $(1-\rho)^N$,
meaning $\rho$ must be very low for very large circuits to complete.
However,
there are scenarios in quantum computing where small subroutines can be performed `offline'
e.g.\ when preparing resource states such as magic states or logical Bell pairs~\cite{Knill2005}.
In these cases, an aggressive repeat-until-success (RUS) model is viable~\cite{Smith2024,Lee2025c}. Apart from the very shallow circuit context of RUS, which has been explored in depth in previous works, we highlight that the less impressive LER improvement by one order of magnitude, achieved at much smaller abort rates, can also be used in some early fault-tolerant applications to reduce resource requirements. 

\subsection{DCS-Based Maximum Likelihood Estimation}\label{sec:MLE}
In this subsection, we explain how MLE can be leveraged in expectation value estimation to improve accuracy.
We call this technique DCS-MLE.

We consider estimating the expectation value of an observable through an experiment  consisting of $M$ noisy runs of the same $N$-window quantum circuit.
Each run $j \in `{1, \dots, M}$ has an observed noisy outcome $z_j \in `{-1, +1}$
and a total LEP $P_{\L,j}$ calculated from its DCS record.
The Methods subsection ``Numerical Simulation of Expectation Value Estimation" details how we emulate these runs.
A straightforward estimator is the unweighted average:
$\langle Z_{\textrm{unmitigated}} \rangle = \frac{1}{M} \sum_{j=1}^{M} z_j.$
However, this can be improved using the data $`{P_{\L,j}}_j$ as follows.

Let us associate a random variable $Z_j$ to each circuit run.
The likelihood to observe the set of outcomes $\{z_j\}_{j=1}^{M}$ is
\begin{equation}
    \mathcal{L} = \prod_{j=1}^{M} \Pr(Z_j = z_j),
\end{equation}
where $\Pr(\text{logical statement})$ is the probability the logical statement is true.
Let the underlying noiseless outcomes
from the logical circuit be $\{\tilde{z}_j\}_{j=1}^{M}$, all having the same mean, $\langle Z \rangle _{\text{th}}$, such that $\theta := \Pr(\tilde{Z}_j=+1)=\frac{1}{2}(1 +\langle Z \rangle _{\text{th}})$.
We can express the probability to observe a particular outcome
in terms of conditional probabilities on the noiseless outcomes:
\begin{equation}
\Pr(Z_j=z_j)
=\sum_{\tilde{z}_j}\Pr(Z_j=z_j|\tilde{Z}_j=\tilde{z}_j)\Pr(\tilde{Z}_j=\tilde{z}_j). 
\end{equation}
Generally, noise corrupts the true outcomes in a non-trivial way,
defined by a corruption function $q$
which depends on the logical error probability: $q(P_{\L,j}, \tilde{z}_j) := \Pr(Z_j=+1|\tilde{Z}_j=\tilde{z}_j)$. Therefore:
\begin{equation}\label{eq:theta_dependence}
    \Pr(Z_j =+1)=\theta q(P_{\L,j}, +1)+(1-\theta)q(P_{\L,j}, -1).
\end{equation}
Finally, noting that $\Pr(Z_j=-1)=1-\Pr(Z_j=+1)$,
the likelihood $\mathcal{L}$
can be written as a function of $\theta$,
with the dataset $`{(z_j, P_{\L, j})}_j$ as known parameters.
We obtain an estimator for $\theta$ which, under the usual regularity conditions, is asymptotically efficient and attains the Cramér–Rao bound by
maximising $\mathcal{L}$,
which in practice is done by minimising the negative log-likelihood,
$-\ln \mathcal{L}$.

The function $q$ will depend on the noise propagation in the particular circuit under consideration.
For simplicity, we assume that any logical error completely randomizes the output, so the noise impact can be modelled as $q(P_{\L,j}, +1) = 1-\frac{1}{2}P_{\L,j}$ and $q(P_{\L,j}, -1)=\frac{1}{2}P_{\L,j}$. Under this assumption, when repeating the expectation value estimation experiment a number $N_{\text{rep}} \to \infty$ of times, the unmitigated estimator distribution for $M$ used shots achieves the following bias and variance: 
\begin{equation}
    \begin{aligned}
\text{Bias}_{\text{unmitigated}} &= -\langle Z \rangle_{\text{th}} \bar{P}_L, \\
\text{Var}_{\text{unmitigated}} &= \frac{1 - \langle Z \rangle_{\text{th}}^2 (1-\bar{P}_\L)^2}{M},
\end{aligned}
\label{eq:unmit_stats}
\end{equation}
where $\bar{P}_\L$ is the mean of the circuit LEP distribution. The maximum likelihood estimator distribution is asymptotically unbiased -- provided that the circuit LEPs are perfectly calibrated and the corruption model \(q\) is correctly specified -- and, from Fisher Information considerations \cite{helstrom_quantum_1969,holevo_probabilistic_2011}, achieves the following variance:
\begin{equation}\label{eq:mle_stats}
    \operatorname{Var}_{\mathrm{MLE}}\approx
    \frac{1}{M}
    \left(
    \mathbb{E}
    \left[
    \frac{(1-P_{\mathrm L})^2}
    {1-\langle Z\rangle_{\mathrm{th}}^2(1-P_{\mathrm L})^2}
    \right]
    \right)^{-1}.
\end{equation}

\subsection{Basic DCS-Agnostic Logical QEM}\label{sec:basic_qem}
In this subsection, we explore whether any progress can be made towards obtaining a better estimate for the observable expectation value in the absence of any DCS knowledge and without applying any other protocols such as those described in the ``Introduction". Under the assumed linear model for the impact of circuit-level logical errors on circuit outcomes, the relation between the unmitigated and the noiseless $(\frac{1}{M} \sum_{j=1}^{M} \tilde{z}_j)$ expectation values is: $\langle Z_{\text{unmitigated}} \rangle=(1-\bar{P}_{\L})\langle Z_{\text{noiseless}} \rangle$. However, in this scenario, we do not have access to $`{P_{\L,j}}_j$; the only additional information available is an estimate of the circuit LER $R\approx\bar{P}_{\L}$ from calibration experiments. Therefore, a natural logical-level QEM strategy is: $\langle Z_{\text{Basic QEM}}\rangle = \frac{\langle Z_{\text{unmitigated}}\rangle}{1-R}$.

We next evaluate the methods from the previous three subsections
on emulated results of an expectation value estimation task
using a distance-11 surface code
(``Distance-11 Surface Code").
Three main observations emerge:
1) all methods significantly lower the estimation error;
2) MLE gives the highest relative improvement over pure abort when applied to unfiltered data;
3) all methods give higher relative improvements
as the minimum number of retained shots increases.
We also construct an approximate model
for the imperfect mapping between $\phi$ and $\lambda$
for a distance-19 surface code
and assess the performance of the abort protocol under this model (``Distance-19 Surface Code").

\subsection{Distance-11 Surface Code}\label{sec:distance_11_surface_code}

We set out to estimate $\langle Z \rangle_{\text{th}}=0.8$ with a $d=11$ surface code,
in a quantum circuit comprising $N\approx 2.57\times10^5$ decoding windows. We consider applications that require at least $M_\text{retained} \in \{\num{1e2}, \num{1e4}\}$ shots to estimate $\langle Z \rangle_{\text{th}}$. The estimation accuracy can be improved by sampling more shots, $M_\text{begun}:=M_\text{retained}/(1-f)$. There are two ways to do this: (1) \textit{information-saving}: complete all $M_\text{begun}$ shots and post-process them at the end using techniques from ``DCS-Based Maximum Likelihood Estimation" and ``Basic DCS-Agnostic Logical QEM"; or (2) \textit{time-saving}: commence the execution of $M_\text{begun}$ shots and abort a fraction $f$ according to ``Abort Protocol", then post-process the $M_\text{retained}$ shots as in case (1).

We consider a range of $f$ values. In case (2), $f$ represents the circuit discard fraction. Therefore, each such value gives rise to a different LEP distribution for the $N$-window circuit.
The Methods subsection ``Numerical Simulation of the Abort Protocol" describes how we generate the datasets, with Figure S3 showing the LER decay against discard fraction -- which makes the abort-only protocol applicable in more general settings, not just expectation value estimation tasks. While the same $f$ values are used for both approaches, the associated multiplicative time costs differ, as detailed under ``Multiplicative Time Cost Caused by QEM" in ``Methods".

\begin{figure*}[tbp]
    \centering
    \includegraphics[width=\textwidth]{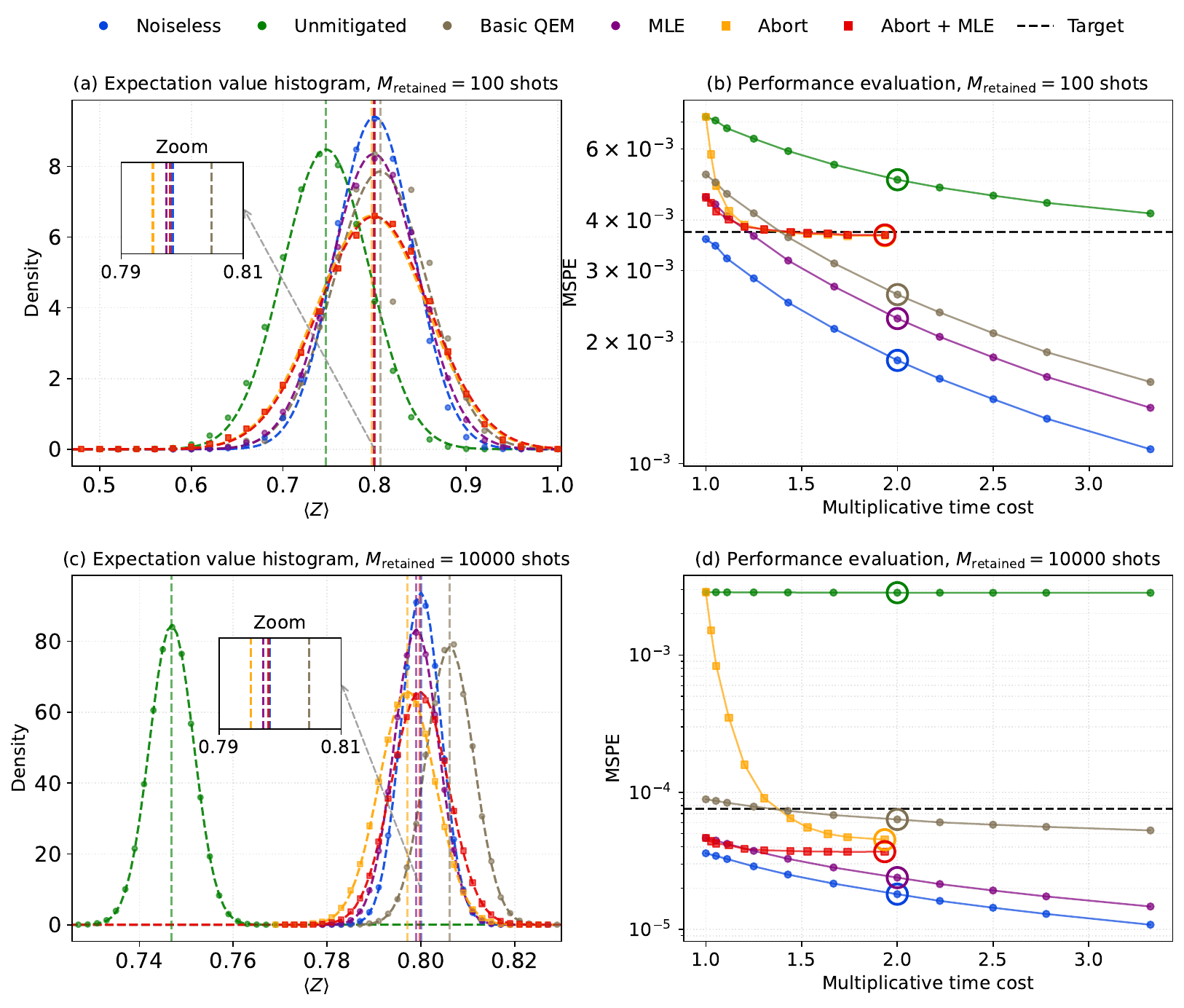}
    \caption{\textbf{Estimating $\langle Z \rangle _{\text{th}}=0.8$ using a distance-11 surface code,
    starting from a logical error rate of $R=7.3(5)\%$ in the absence of error mitigation, learned from calibration data.}
    (a,c) Output distributions of six expectation value estimators, obtained from $N_{\text{rep}}=2\times10^5$ repetitions, with a fixed minimum number of retained shots $M_\text{retained}$.
    Dashed curves are normal distributions with the same mean and variance as their respective datasets,
    which are shown as dots.
    Orange and red datasets estimate $\langle Z \rangle$ from $M_\text{retained}$ shots, after discarding a fraction $f=70\%$ of the total attempted executions. The other four datasets estimate $\langle Z \rangle$ from $M_\text{begun}=M_\text{retained}/(1-f)$ shots, $f=50\%$, with no executions discarded. The noiseless estimator (blue) corresponds to a noiseless quantum computer
    whose outcome variation is caused solely by statistical Bernoulli fluctuations. Basic quantum error mitigation (QEM; grey) and Maximum Likelihood Estimation (MLE; purple) are applied to the unmitigated dataset (green).
    (b,d) The performance of the six estimators plotted against time overhead. MSPE is the mean squared prediction error,
    defined in \cref{eq:mspe_definition}. The `target' MSPE is specific to the Hubbard model application described in ``Application: Hubbard Model Resource Estimation" and it is calculated for $M_\text{retained}$ shots, thus independent of $f$.
    The multiplicative time cost represents how much longer the quantum computation takes compared to commencing and fully executing $M_\text{retained}$ shots; its calculation is explained in ``Multiplicative Time Cost Caused by QEM". Circled data points measure the MSPE of the distributions plotted in panels a,c.}\label{fig:MLE_abort}
\end{figure*}

In \cref{fig:MLE_abort}a,c, we show the distribution of outputs from six different estimators, obtained from $N_{\text{rep}}=2\times10^5$ repetitions of the expectation value estimation experiment.
\Cref{fig:MLE_abort}b,d shows how they perform as the time cost varies.
We choose the mean squared prediction error (MSPE):
\begin{equation}\label{eq:mspe_definition}
    \MSPE :=\frac{1}{N_{\text{rep}}}
    \sum_{i=1}^{N_{\text{rep}}}|\langle Z \rangle_i-\langle Z \rangle_{\text{th}}|^2
\end{equation}
as a performance metric, since it penalises both bias and variance in the estimated expectation values ($\MSPE=\text{Bias}^2+\text{Variance}$).
Other metrics could be selected based on the task at hand, such as the absolute mean bias
(explored in Supplementary Results)
or the probability to obtain an estimate within a desired range.

There are three effects to be noted. 

First, the performance of all methods improves monotonically as the time cost increases.

Second, the benefits of all methods are more pronounced when the minimum number of retained shots $M_\text{retained}$ increases, because statistical fluctuations are averaged out and variances are reduced. For $M_\text{retained}=10^4$, given a set of shots (filtered or unfiltered), further applying MLE is the most effective mitigation protocol. 

Third, with respect to our chosen range of $f$ values, abort-based protocols show diminishing returns. In the case of abort-only (orange), \cref{eq:unmit_stats} predicts that the MSPE scales with the squared retained mean LEP $P_{\text{kept}}^2$. The latter requires an exponential sampling overhead for exponential suppression, as subsequently found in \cite{tsubouchi_quantum_2026}. For MLE applied on top of aborting (red), there are two competing factors as the filter gets more aggressive: on the one hand, the retained shots are cleaner and thus more reliable; on the other hand, their range of values narrows down, so they become less informative.

Finally, for QEM protocols applied on unfiltered data (grey and purple), the MSPE scales inversely with the total number of shots $M_\text{begun}$, akin to the noiseless estimator. Therefore, the time-saving strategy is preferable when only small time costs are tolerable (the MSPE of Abort+MLE is lower than that of MLE-only for a small range of $f$ values), but the information-saving strategy performs better if significant time costs are allowed (e.g. above $50\%$).

In the above discussion, to investigate the abort protocol, we fixed the number of retained shots $M_\text{retained}$ and allowed for an increase in the total number of circuit executions $M_\text{begun}>M_\text{retained}$. Supplementary Results presents the complementary angle: fixed runtime budget and variable number of retained shots depending on the discard fraction.

\subsection{Distance-19 Surface Code}\label{sec:distance_19_surface_code}
We would also like to evaluate the performance of our proposed methods in the context of a realistic application.
The specific algorithm we consider
(discussed in detail under ``Application: Hubbard Model Resource Estimation")
imposes a maximum allowed single-window LER of $r_{\max} =\num{1e-11}$.
Given a physical error rate $p_\text{phys} =\num{1e-3}$,
the smallest code distance that satisfies this is 21, and we aim to reduce it to 19 using DCS-based QEM.

\begin{figure}[tbp]
    \centering
    \includegraphics[width=1.0\linewidth]{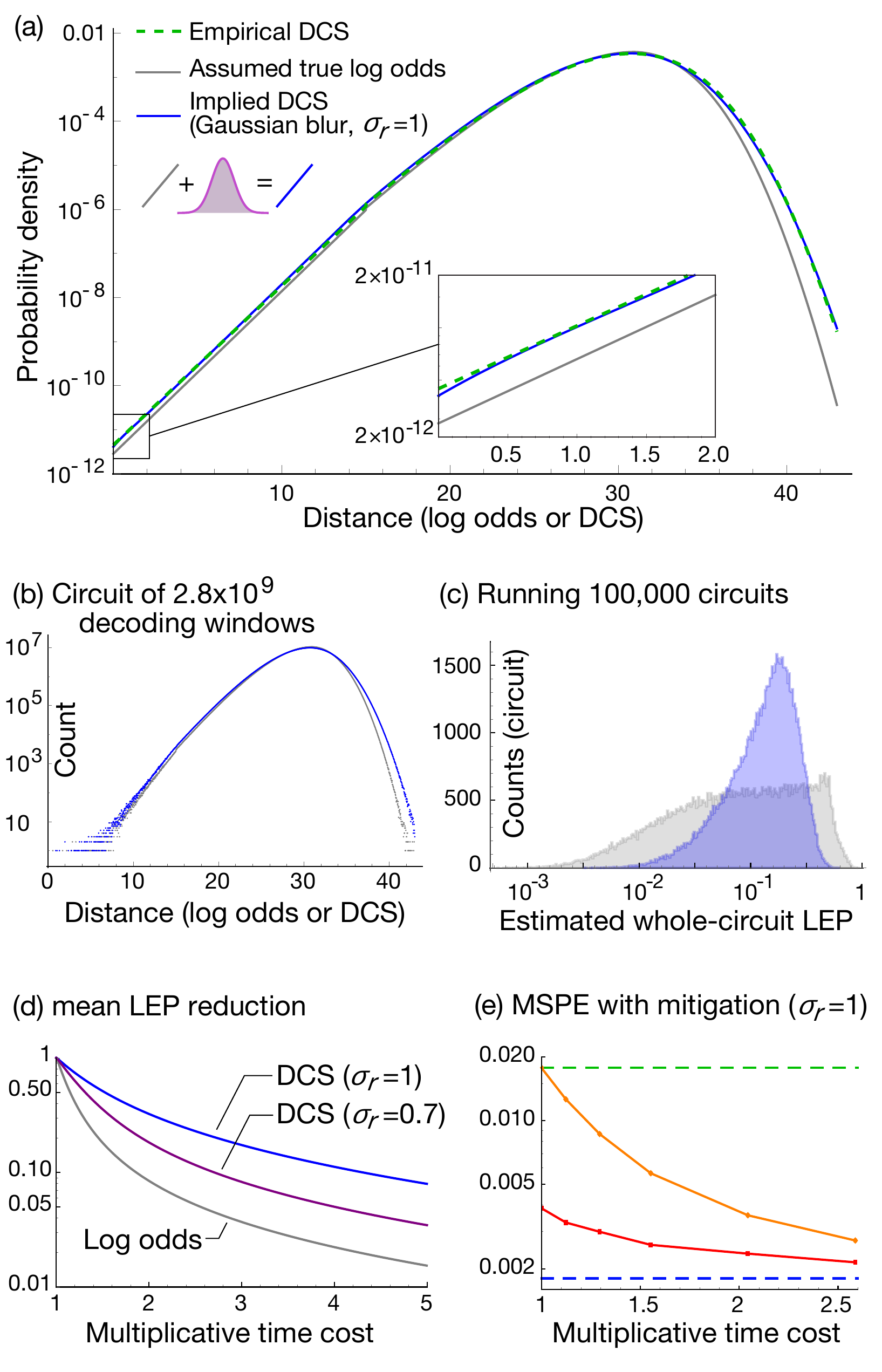}
    \caption{\textbf{Contrasting the at-scale performance of the decoder confidence score (DCS; blue)
    versus the ideal but impractical-to-compute log odds $\lambda$ (grey).}
    In all cases the chosen DCS is the swim distance.
    In panel (a) the green dashed line shows the fitted probability density function of \emph{measured} DCS values for decoding windows of a distance-19 surface code, by binning the calibration data into 500 bins.
    An assumed probability curve for $\lambda$ is shown in grey, and in blue an \emph{implied} DCS distribution derived by redistributing $\lambda$
    via a Gaussian with standard deviation $\sigma_r =1$.
    We see a good match between green and blue.
    Using the distributions in (a) we generate sets of \num{2.8e9} scores, representing all decoding windows in a large circuit; an example distribution is shown in (b).
    From such a set we infer the logical error probability (LEP) $P_\L$ in the circuit; repeating this 10$^5$ times we obtain distribution (c) with mean $P_\L=0.153$.
    We note that relying on the DCS means we cannot identify very high or very low risk circuits, as we would with access to $\lambda$.
    Panel (d) shows the factor by which we reduce error rates using abort criteria, at the cost of increased quantum processor time.
    Panel (e) is the equivalent of \cref{fig:MLE_abort}b,d with the same meaning for lines,
    but with an intermediate number of shots: 200.}\label{fig:d=19model}
\end{figure}

Note that, regardless of $p_\text{phys}$,
we would need to emulate an impractical number $\sim1/r_{\max}$ of single-window memory experiments
to benchmark our methods using Monte-Carlo sampling.
We therefore instead estimate its performance with the following model.
In the absence of a known distribution of the log success odds $\lambda$,
we use the distribution of measured DCS values for $d=19$ and slightly deform it, until the mean $p_\L$ associated with this deformed distribution matches
the one predicted from our statistics (see Figure S4a). The Methods subsection ``Modelling the Distance-19 Surface Code" explains how we performed this deformation.
We then use \cref{eq:dcs_as_sum_of_2_rvs}
and assume $r$ follows a Gaussian distribution with mean zero and standard deviation $\sigma_r=1$.
The inferred distribution of DCS values closely matches that of the measured ones, as seen in \cref{fig:d=19model}a.
This is therefore an entirely analytic model,
in which we have descriptions of $\lambda$ and the DCS.
Moreover, we know the range of $\lambda$ associated with each DCS value,
so there is no need to calibrate the DCS --
in a real scenario, imperfect calibration would indeed impact performance;
further work could explore the impact of this.

\Cref{fig:d=19model}d shows
impressive improvement in the whole-circuit LEP when we impose an abort condition based on the instantaneous window DCS.
As suggested by the observation in ``Many-Decoding-Window Statistics"
that most of the logical error risk is contained in a few very risky windows,
we see the post-selection method does not require exponential overhead for a linear decrease in LER (for a significant range of reduction).
However, we do observe that the amount of achievable reduction is very sensitive to DCS accuracy.
Recall that the results under ``Decoder Confidence Scores are an Excellent Indicator of Logical Error Risk" established a close match between the swim distance and $\lambda$;
even so, we see in \cref{fig:d=19model}c,d that the imperfect correlation manifests in a considerable worsening of the quality of the information we obtain about the whole-circuit LEP.
Note that the $\sigma_r=1$ assumption is somewhat pessimistic as an extrapolation from \cref{fig:source_of_dcs_variation_perturb0.3},
and with a modest reduction to $\sigma_r=0.7$ the performance dramatically improves.
Conversely, if we were to replace the swim distance with an alternative DCS that is significantly weaker,
it is possible that the whole-circuit LEP mitigations explored in this paper would become ineffective.

Finally, in \cref{fig:d=19model}e we generate
the equivalent of the earlier panels in \cref{fig:MLE_abort}b,d.
It is reassuring that the key qualitative features are again seen in this alternative model, despite the significant $\sigma_r$ value. The reduction in whole-circuit LEP variance and the indistinguishability of high-risk shots reduce the amount of information available to DCS-MLE, thus diminishing its performance. A similar effect is produced by additional imperfections, such as calibration errors (which change the edge weights in the decoding graph) and noise model mismatch (which changes the structure of the decoding graph). These could potentially introduce systematic calibration bias, which is an additional failure mode not captured by our phenomenological sensitivity model for zero-mean uncertainty.

\subsection{Application: Hubbard Model Resource Estimation}\label{sec:res_est}
We now discuss the application of the abort protocol and DCS-MLE in a quantum phase estimation (QPE) algorithm. We apply our findings to reduce the resource requirements for the implementation of a full logical circuit.
Specifically,
we find that the surface code distance can be reduced from 21 to 19
while achieving the same LER with a $44\%$ time overhead via the abort-only protocol, or the same MSPE with a $25\%$ time overhead via MLE-only. 

We target estimating the ground state energy of the $10 \times 10$ Hubbard model to $10^{-2}$ statistical accuracy (in the absence of physical noise)
and at most $10^{-2}$ LER (from physical noise added on top of statistical fluctuations),
following the QCELS-based~\cite{ding_even_2023} resource estimates in Ref.~\cite{akahoshi_compilation_2024}.
The algorithm involves executing $N_{\text{circuits}}=160$ different quantum circuits of various depths,
each repeated with $M_\text{retained} = 100$ shots to estimate an observable's expectation value. This means that $N_{\text{executions}}=N_{\text{circuits}}M_\text{retained}$ circuit executions would take place in practice. 
The Methods subsection ``QCELS Parameters" provides more algorithmic details. We estimate that an implementation that does not use the DCS
requires a $d_\text{targ} =21$ surface code, which corresponds to a LER of approximately $R_\text{targ} =8.3\times10^{-3}$;
this is our target LER.

The potential advantage of the DCS arises during the estimation of expectation values.
We explore whether the necessary quantum circuits can be executed with a smaller $d_\text{small} =19$ code
[at an unmitigated LER of $R_\text{small} =7.3(5)\%$]
and still meet the target LER via QEM.
Therefore, the desired improvement in the LER is
$R_\text{small}/R_\text{targ}=8.75$.
We emulate the behaviour of the $d=d_\text{small}$ code
using $d=11$ and \cref{rem:rescale_distributions}.
Under ``Distance-11 Surface Code",
the number of decoding windows was chosen precisely to approximate the LEP distributions for $d=d_\text{small}$.

First, we investigate leveraging the DCS via the safest method, namely the abort protocol, because it does not require any assumptions about the impact of noise on logical outcomes. By repeatedly sampling $N_{\text{executions}}$ values from the emulated $N$-window circuit output distribution used under ``Distance-11 Surface Code" at a $f=50\%$ discard fraction,
we estimate that the 95\% confidence interval for the effective LER is $[5.2, 7.7]\times10^{-3}$,
comfortably below the target LER,
at a $44\%$ time overhead and overall $7\%$ increase in space-time volume (calculated under ``Code Distance Influence on Resources" in ``Methods").

Next, we study the DCS-MLE benefits. For this algorithm,
it may be more useful to analyse
the MSPE instead of the effective LER.
Using the data shown in \cref{fig:MLE_abort}b, we find that the MSPE specific to the target code distance
is achieved by the MLE-only strategy at a lower time overhead of $25\%$ and an overall $7\%$ decrease in space-time volume. 

To summarise,
we explore the Hubbard model application in a scaled-down form
which allows us to apply our earlier empirical results from the ``Distance-11 Surface Code" subsection.
We believe that for the full problem
it should be possible to reduce the surface code distance from $21$ to $19$ with only a modest change in space-time volume (slightly positive or negative, depending on the chosen QEM strategy).
The advantage here is that the application could run successfully on a smaller quantum computer.

\section{Discussion}\label{sec:discussion}
Concurrent with our work,
Ref.~\cite{Zhou2025b} proposes
similar ideas on using the DCS for QEM.
For example,
the authors consider using the DCS for entire logical circuits
and reach similar conclusions using the same abort-based method;
they also assess the accuracy of the complementary gap
using a tensor network decoder.
While similar,
there are differences.
The authors use a code capacity noise model
(i.e.\ assume noiseless measurements)
and emulate logical gates;
we adopt a more realistic circuit-level noise model
but emulate only quantum memory.
The authors propose a ZNE procedure based on probabilistic error amplification,
whereas the MLE procedure we propose is based on the LEP variation between runs.

Shortly after our preprint,
two related preprints were announced. Ref.~\cite{aharonov_syndrome_2025} proposed a logical QEM protocol called SALEM, conditioned directly on the decoder syndrome. The authors treat different syndromes as inducing different logical error channels, and syndrome-dependent mitigation or rejection strategies are applied to reduce logical errors at the channel level. While both our work and theirs exploit syndrome-derived run-to-run LEP variability and may reject high-risk executions, they differ substantially in methodology. For accepted executions, SALEM mitigates errors via inverting logical error channels, whereas our approach further uses the DCS as a parameter in classical post-processing via maximum-likelihood estimation. The requirements to implement the methods are also different: SALEM requires channel characterisation, while DCS-MLE requires a calibrated mapping between the window-level DCS and the LEP.
The second preprint, ref.~\cite{tsubouchi_quantum_2026}, analysed the asymptotic logical error rate improvement achievable via similar QEM methods using an information-theoretic framework, and explored the possibility of adapting the logical quantum control based on the observed syndrome.

This paper brings more nuance to the existing works on decoder confidence
and provides a proof of principle for error mitigation. 
We identify four limitations/further directions of research.

Firstly,
the tensor network method we used to compute $\lambda$
was prone to numerical imprecision
as there were many samples that led to
division by zero or negative odds.
For the remaining samples,
floating-point precision capped the $\lambda$ values at 16,
meaning we were analysing a very restricted subset of cases,
especially at large code distances, where it is known that
$\lambda$ can far exceed 16.

Secondly,
to maintain decoding accuracy,
most decoding schemes~\cite{Dennis2002,Skoric2023,Bombin2023,Chan2026}
require neighbouring decoding windows in a quantum computation to overlap:
for a code of distance $d$,
the overlap is usually of the order $d/2$~\cite{Lin2025a}.
This means decoding windows are not independent;
accounting for this effect would alter \cref{ler_many_windows}.
Moreover, our numerical results for $N$-window circuits made use of emulated logical errors, obtained under the assumption of window independence,
so they should be trusted with caution.

\begin{figure*}[tbp]
    \centering
    \includegraphics[width=\linewidth]{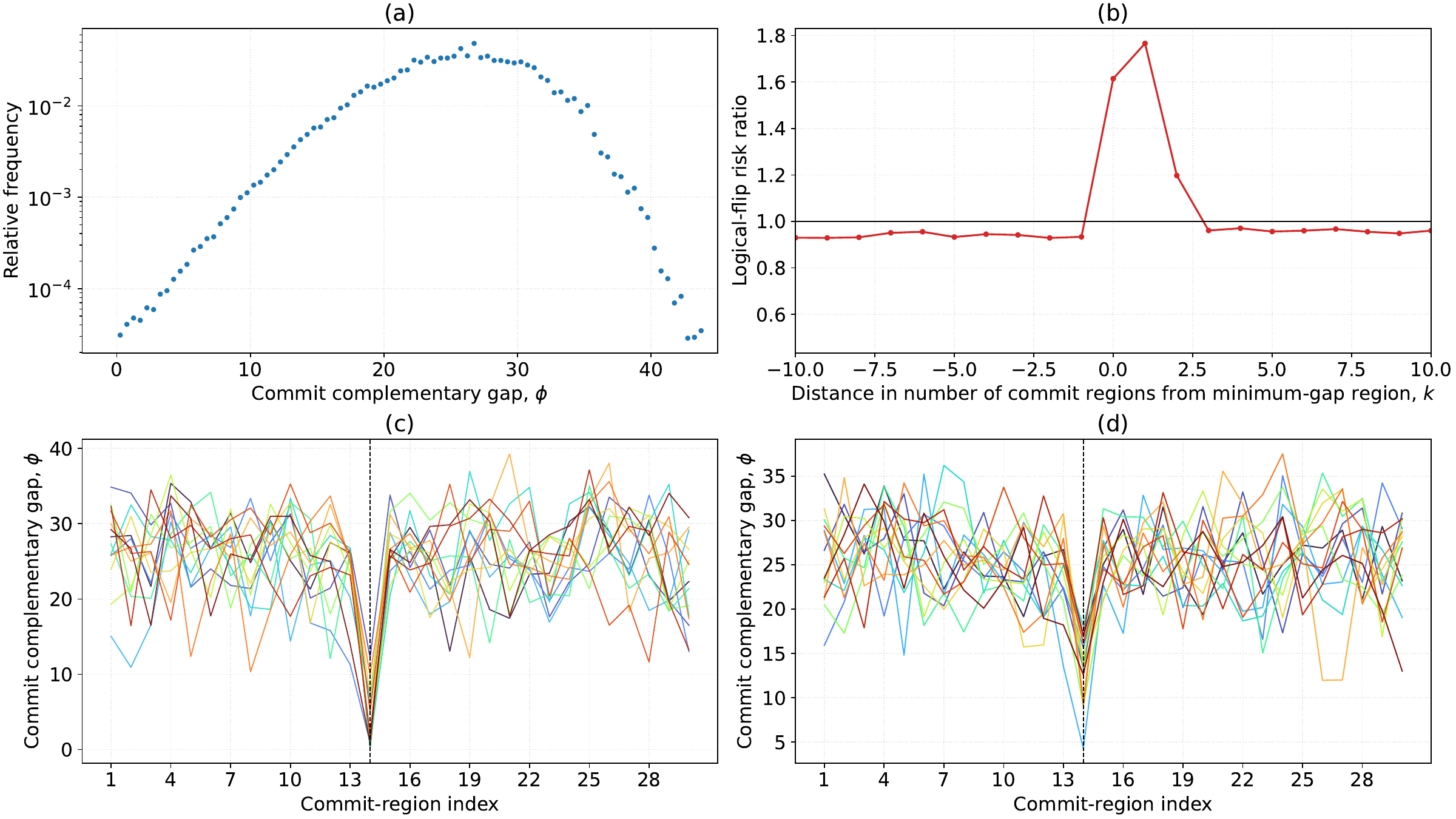}
    \caption{
    \textbf{Locality of correlations and DCS information in sliding-window decoding.}   
    (a) Distribution of the decoder confidence score (DCS) $\phi$ over all commit-region samples, shown using
    88 equal-width bins.
    (b) Localization of logical risk around the minimum-gap region. For each shot, commit regions are aligned relative to the region
    $i_{\min}$ containing the minimum $\phi$ of the shot; the plotted quantity is the
    local logical-flip probability at displacement $k=i-i_{\min}$,
    normalised by the unconditional local flip probability: $P(\delta_i=1)/P(\delta=1)$, where $\delta_i$
    denotes the logical-parity increment associated with commit region $i$.
    Logical risk is enhanced in the minimum-gap region and the immediately
    following region and rapidly returns towards its baseline value away
    from this neighbourhood.
    (c) DCS records for all 14 shots with a final
    logical error whose first minimum $\phi$ occurs at commit region 14. Each colour corresponds to a different shot.
    (d) DCS records for 14 randomly selected successfully decoded
    shots whose first minimum $\phi$ also occurs at commit region 14, sampled from
    3342 qualifying shots.
    The dashed vertical lines in (c,d) mark commit region 14.}\label{fig:window_indep}
\end{figure*}

In a sliding-window decoder, the artificial
defects produced by committed corrections are propagated to subsequent
windows. To assess the practical implications of these correlations, we performed
an additional circuit-level simulation of a rotated distance-$11$
surface-code memory at $p_{\mathrm{phys}}=2.5\times10^{-3}$, using
$10^5$ shots, $30$ sequential $d$-cycle commit regions, and a $d$-cycle
overlapping buffer. For each commit region, we evaluated a spatial
complementary-gap DCS, $\phi$, while retaining the buffer and temporal-boundary
parities selected by the sequential decoder. Its distribution closely resembles the shape of globally-decoded DCSs, as shown in \cref{fig:window_indep}a.

Our results suggest that the resulting logical correlations are
strong, but highly localized. Conditional on a logical-parity flip in one
commit region, the probability of another flip in the immediately following
region is enhanced by approximately a factor of $15.5$ relative to the
unconditional flip rate. At a separation of two commit regions, the risk ratio
already returns to approximately unity and remains close to unity at larger
separations. 

As shown in \cref{fig:window_indep}b, the DCS exhibits a correspondingly local relation to logical risk.
Aligning individual executions by the commit region containing their minimum
$\phi$, the logical-parity-flip probability is enhanced in the minimum-gap
region and the immediately subsequent region, but rapidly returns towards its
unconditional value away from this neighbourhood. The individual DCS records in
\cref{fig:window_indep}c,d illustrate this behaviour:
among shots conditioned to have their DCS minimum at the same commit region,
executions ending in a logical error typically exhibit substantially deeper
troughs than successfully decoded executions.

These results show that the independent-window assumption should be regarded
as a simplifying zeroth-order model rather than an exact property of
sliding-window decoding. Importantly, however, the correlations observed here
have a short temporal range, and low DCS values identify the neighbourhood in
which the enhanced logical risk is concentrated. This suggests that
correlation-aware extensions may only need to retain local temporal information, for
example by jointly calibrating a learning agent over a small number of adjacent windows or by
selectively redecoding a low-confidence neighbourhood using a larger temporal
window, as suggested by the double window decoding scheme in Ref. \cite{Toshio2025}. We leave a systematic investigation of such correlated extensions to future work.

Thirdly,
regarding DCS-based QEM, the calibration curves for large code distances take a long time to obtain because the LEPs are so low. Thus, full calibration curves would have to be provided by the quantum device manufacturers. Nevertheless, \cref{fig:calibration_curve}b shows that the calibration curves are extremely similar for the same noise model between different code distances, so an approximate calibration curve for $d=19$ could be derived from lower distances.

Lastly,
we emulated deep logical circuits by stitching together the results from individual memory experiments. A full fault-tolerant treatment would be more involved, but, in principle, feasible. 
For example, it would have to account for multiple interacting logical qubits,
meaning the corruption function $q$ would in general be nonlinear in $P_{\L, j}$. What is needed to even calculate $P_{\L, j}$ is an efficient multi-qubit decoder and a well-calibrated DCS for the logical qubit interaction. In the case of the surface code, the interaction is performed via lattice surgery, and ref. \cite{Akahoshi2025} suggests a temporal complementary gap to estimate the probability of a measurement error. However, this is not the only possible type of error, and an overall efficient-to-compute DCS remains a future research direction. Given the complexity of lattice surgery decoding, we expect that our approach slightly underestimates the effective number of decoding windows required by a given application and thus overestimates the achievable error rates, but the impact would not be profound.
Note that a logical Pauli noise model can be constructed for general circuits based on the DCS.
We can consider a computational model with layers of logical operations, each of which lasts for a multiple of $d$ stabiliser measurement cycles. Hence, each non-interacting logical qubit within a layer experiences an $X$, $Y$, or $Z$ error, while pairs of interacting logical qubits likewise experience a weight-two Pauli error. The probabilities of these errors are estimated from the DCS of each decoding instance and type. 

A fundamental question concerning our proposed methods is whether the abort protocol is scalable to large code distances (beyond the early fault-tolerant era). Most existing works have focused on
applications of the DCS with small (\num{<11}) distances
due to scalability:
the sampling overhead of the abort protocol could be unreasonably high.
The results under ``Many-Decoding-Window Statistics" provide numerical evidence against this idea by considering the statistical properties of the DCS distribution
and demonstrating that the method works for a small code with the same LER as a larger one. Even though exponential error suppression still requires exponential sampling overhead \cite{tsubouchi_quantum_2026}, gaining a factor of $10$ in the error rate at the cost of order $50\%$ time overhead could nevertheless be valuable under restrictive qubit counts.
Ref.~\cite{Smith2024} provides alternative evidence towards scalability:
the authors find numerically that
the rate at which the decoder aborts a circuit run
exhibits threshold behaviour.
Namely, for a fixed abort tolerance,
if the physical error rate is below (above) some threshold,
the abort rate decreases (increases) as the code distance increases.
They observe that the threshold is on the order of $10^{-2}$ for the code capacity noise model, and expect it to decrease under more realistic noise models. Nevertheless, as long as the abort threshold does not drop below $10^{-4}$, post-selection remains a valuable tool for useful applications in the early fault-tolerant era.

Although we have focused on MWPM decoding of the surface code and estimated logical risk using the swim distance, our methods for full-circuit logical error mitigation should apply more generally provided two conditions are satisfied: (1) the decoder outputs a DCS that is well calibrated to the true log odds (a decoder property); and (2) the distribution of full-circuit LEP values remains sufficiently broad after aggregating many decoding windows (a code property). For example, DCS-informed post-selection for qLDPC codes has recently attracted considerable attention. Confidence metrics based on cluster statistics \cite{Lee2025c}, the logical observable flip probability predicted by neural-network decoders \cite{gu_scalable_2026, Dentelski2026}, and the forced gap \cite{wills_forced_2026} have all been shown to reduce logical error rates at modest discard fractions for shallow circuits (up to 15 syndrome extraction cycles). With respect to our framework, Supplementary Note 1 of Ref.~\cite{Lee2025c} suggests that their proposed DCS does not satisfy condition (1), as its correlation with the true gap is weak. By contrast, the DCS constructions in the other works provide stronger evidence that condition (1) is met. In Ref.~\cite{gu_scalable_2026}, the decoder explicitly estimates the probabilities of the different logical classes and bases its decoding decision on these estimates, making accurate LEP knowledge a prerequisite for good decoding performance. In Ref.~\cite{wills_forced_2026}, the forced gap is a practical extension of the surface code complementary gap, so we expect it to exhibit similar calibration properties, but further work is necessary to quantify this. Whether these methods also satisfy condition (2) remains an open question.

We began by showing that the log success odds of a decoding instance can be very well approximated by both the complementary gap and the swim distance for MWPM under phenomenological noise,
justifying their status as `decoder confidence scores' (DCSs).
For a simplified model of window-based decoding under circuit-level noise,
we have shown that DCSs remain rich in information as the circuit size increases, which suggests that they can be used for more complex tasks than state preparation. Finally, we have explored this latter idea in the form of QEM techniques for the early fault-tolerant quantum computing era, finding that, within our emulated independent-window model, one can accurately estimate the logical error risk of an entire circuit and leverage this information with maximum likelihood methods.
We emphasise that our results rely on the observation that the swim distance is an excellent DCS; the performance of both proposed QEM methods
diminishes as the DCS quality decreases.
This motivates future work toward even more accurate DCS methods.

\section{Methods}
\subsection{Calculating the Sources of Variation of the DCS}\label{sec:calculating_the_sources_of_variation_of_the_dcs}
First, we describe how we obtained $\sigma_r$.
We emulated the unrotated surface code
under a perturbed phenomenological noise model
(described under ``Perturbed Phenomenological Noise Model" in ``Methods").
For odd distances \numrange{3}{11},
we randomly sampled syndromes obtained from noise levels in \numrange{1e{-4}}{1e{-1}}.
For each syndrome,
we calculated both DCS values using the MWPM decoder~\cite{Higgott2025},
then calculated $\lambda$ via tensor-network contraction~\cite{piveteau_tensor-network_2024}
(details under ``Tensor Network Contraction").
The results are shown in Figure S1
and suggest that $\alpha$ is linear for both DCSs.
Therefore,
we fit a line $\hat{\phi}(\lambda)$ to the data $`{(\lambda_i, \phi_i)}_{i =1}^n$ for each distance
and estimate $\sigma_r$ by calculating the \emph{sample root mean squared residual}:
\begin{equation}\label{eq:sample_root_mean_squared_residual}
    \widehat{\sigma_r} =
    s_r :=\sqrt{\frac{\sum_{i =1}^n `\big[\phi_i -\hat{\phi}(\lambda_i)]^2}{n -1}}.
\end{equation}

To obtain $\sigma_{\alpha(\lambda)}$,
we calculated both DCS values for \num{1e8} randomly sampled syndromes
at a fixed noise level of \num{1e-3},
then calculated the sample variance $s_\phi^2$ for each DCS.
Finally,
we note that \cref{eq:dcs_as_sum_of_2_rvs}
and the independence of $\lambda$ and $r$ implies
$\sigma_\phi^2 =\sigma_{\alpha(\lambda)}^2 +\sigma_r^2$,
so we estimate $\sigma_{\alpha(\lambda)}$ by calculating
\begin{equation}
    \widehat{\sigma_{\alpha(\lambda)}} =\sqrt{s_\phi^2 -s_r^2}
\end{equation}
where $s_r$ is taken from \cref{eq:sample_root_mean_squared_residual}.

\subsection{Perturbed Phenomenological Noise Model}\label{sec:perturbed_phenomenological_noise_model}

In the phenomenological noise model,
all edges have the same weight,
giving rise to a very limited set of possible values the DCS can take.
We found that this led to a large value for $s_r$ in \cref{eq:sample_root_mean_squared_residual};
however,
such a simplistic noise model is unrealistic.
Ideally,
we would perform our analysis under ``Decoder Confidence Scores are an Excellent Indicator of Logical Error Risk"
using a circuit-level noise model
whose edge weights are calibrated from experiments or existing syndrome data.
Although Ref.~\cite{piveteau_tensor-network_2024} provides a tensor-network decoder
for circuit-level noise,
we could not obtain values of $\lambda$ that converged quickly enough.
Therefore,
we resorted to a `perturbed' phenomenological noise model,
wherein each edge weight is multiplied by a random scalar drawn uniformly from $[1 -\delta, 1 +\delta]$.
We set $\delta =0.3$ which roughly corresponds to the variation in edge weights
for standard circuit-level depolarising noise of strength $p =\num{1e-3}$
\cite{Gidney2021a}.

\subsection{Tensor Network Contraction}
\label{sec:tensor_network_contraction}
We use the software package \verb|tndecoder3d| introduced in Ref.~\cite{piveteau_tensor-network_2024}.
For the surface code under phenomenological noise,
the network to contract resembles a 3D cubic lattice.
The contraction algorithm first contracts the 3D lattice down to a 2D lattice,
then contracts the 2D lattice using well-known techniques.
Following Ref.~\cite{piveteau_tensor-network_2024},
we use a maximum bond dimension of 24 and 32 for each stage respectively.
Figure S2 shows that
these values, respectively named `3D bond dimension' and `2D bond dimension',
are sufficient for $s_r$ in \cref{eq:sample_root_mean_squared_residual} to converge.

\subsection{DCS-Log Success Odds Calibration}\label{sec:calib}
The calibration being treated as a linear function is an empirical observation for the studied noise model and the chosen DCSs, an observation which has also been made in previous works for the complementary gap \cite{Gidney2025, Akahoshi2025}, the latter noting that this behaviour is yet unexplained. Ref. \cite{meister_efficient_2024} proved that the swim distance is exactly equal to the log success odds for the repetition code, and that it is at most equal to the complementary gap for the surface code. We can begin to theoretically understand the behaviour of the complementary gap calibration by writing out all the terms contributing to the log success odds. Let us denote the logical equivalence class of the MWPM solution as $\text{LEC}_1$, and the opposite logical equivalence class as $\text{LEC}_2$. An error chain belonging to a LEC will be denoted by $C$. $p_{\text{max},i}$ represents the probability of the most likely error chain within $\text{LEC}_i$; $p_{\text{rest},i}$ is the total probability of all other error chains belonging to that LEC. The log success odds are thus:
\begin{equation}
\begin{aligned}
\lambda &:= \log_{10} \frac{\Pr(\text{LEC}_1)}{\Pr(\text{LEC}_2)} \\
&= \log_{10} \frac{\sum_{C \in \text{LEC}_1} \Pr(C)}{\sum_{C \in \text{LEC}_2} \Pr(C)} \\
&= \log_{10} \frac{p_{\max,1} + p_{\text{rest},1}}{p_{\max,2} + p_{\text{rest},2}} \\
&= \underbrace{\log_{10} \frac{p_{\max,1}}{p_{\max,2}}}_{\phi} + \log_{10} \frac{1 + \frac{p_{\text{rest},1}}{p_{\max,1}}}{1 + \frac{p_{\text{rest},2}}{p_{\max,2}}} \\
&= \phi + \log_{10} \frac{1 + \frac{p_{\text{rest},1}}{p_{\max,1}}}{1 + \frac{p_{\text{rest},2}}{p_{\max,2}}.}
\end{aligned}
\end{equation}
The probability of a LEC is dominated by its most likely error chain ($p_{\text{rest,}i}\ll p_{\max,i}$), so, to zeroth order, the log success odds $\lambda$ are equal to the complementary gap $\phi$. The expansion of the remaining fraction is tricky, because the division of LEC probability between the likeliest error chain and all the others varies with $\phi$. Nevertheless, if this correction varies slowly with $\phi$, or similarly between the two LECs, an approximately linear calibration can emerge.

We note that other monotonic, non-linear functions could be needed, depending on the code, noise model, decoder, and DCS choice.

\subsection{Numerical Simulation of the Abort Protocol}\label{sec:abort_simulation}\label{sec:num_sim}
\setcounter{figure}{0}
Here, we describe how we generate datasets for the LEP and logical error of each $N$-window circuit.
We perform simulations with a code distance of $11$,
which is the largest distance at which enough logical errors could be detected during calibration within the finite number of shots to ensure representative large-circuit statistics. To avoid simulating deep circuits and calculating their DCS values,
we emulate the behaviour of a circuit comprising $N =\num{2.57e5}$ windows by sampling from a pool of single-window circuits, each characterised by a calibrated LEP $p_{\L,i}$ and a logical error $x_i \in \{0,1\}$, where $x_i = 0$ indicates successful decoding. The $N$-window circuit is considered to have experienced an error if the decoding of an odd number of the $N$ windows failed: $X = 1$ if $\sum_{i=1}^Nx_i$ is odd, else $X=0$. The assigned LEP for an $N$-window circuit is given by \cref{ler_many_windows}.

Next, we explain how the abort protocol affects the LEP distribution. Let the single-window abort rate be $\rho$, achieved by setting some LEP threshold. This means that, in an $N$-window circuit, each window has an LEP larger than that threshold with probability $\rho$. The probability to discard the circuit by aborting at any point throughout its run is $f=1-(1-\rho)^N$. Throughout this work, we refer to $f$ as the discard fraction. This is analogous to \cref{fig:d=19model}d.

To obtain the LEP distribution for a certain abort rate, we eliminate all the single-window experiments from the $10^9$ cohort whose LEP is above the threshold LEP. We then proceed as described in the first paragraph of this subsection and sample $N$ windows $5\times10^6$ times. In Figure S3, we show the retained LER decay with discard fraction.

For the resource estimation task described in ``Application: Hubbard Model Resource Estimation", we notice that a $f=50\%$ discard fraction brings the LER down to the target value. Therefore, we sample $N_{\text{executions}} = \num{16000}$ times from the $\{(X_j, P_{\L,j})\}$ dataset obtained at the abort rate associated with this discard fraction, effectively emulating $N_{\text{executions}}$ instances of $N$-window circuits.
The empirical LER is the mean $X$ across the $N_{\text{executions}}$ accepted shots,
and is stated in the main text with a $95\%$ confidence interval.

\subsection{Numerical Simulation of Expectation Value Estimation}\label{sec:dcs_zne_simulated_data}
Here, we describe how we generate datasets to emulate $M$ noisy shots, each being an $N$-window circuit execution. First,
we emulate $M$ copies of a $\ket{\overline{0}}$ memory experiment as described in ``Numerical Simulation of the Abort Protocol"
to obtain the LEPs $`{P_{\L,j}}_{j=1}^M$ from the DCS
and the logical errors $`{X_j}_{j=1}^M$.
Next, for a fixed ideal expectation value $\langle Z\rangle_\text{th}$,
we employ a random number generator to generate the noiseless logical outcomes $\{\tilde{Z}_j \in \{-1, +1\} \}_{j=1}^M$, where $\tilde{Z}_j=+1$ occurs with probability $(1+\langle Z\rangle_\text{th})/2$.
We then randomise the signs of the noiseless outcomes according to the logical errors $\{X_j\}_{j=1}^M$ to obtain the noisy outcomes $Z_j=\tilde{Z}_j(1-X_j)-UX_j$, where $U$ is a random variable with equal probability to take value $+1$ or $-1$. This means that the noise corruption function from ``DCS-Based Maximum Likelihood Estimation" takes the form: $q(P_{\L,j}, +1) = 1-P_{\L,j}/2$ and $q(P_{\L,j}, -1)=P_{\L,j}/2$.
This setup provides an effective framework for modelling arbitrary logical circuits similar to memory experiments,
both when using and not using the abort protocol. Note that we had to make an assumption about how the noise influences the final outcome. For a memory experiment, the outcome would actually be $100\%$ flipped by a logical error, but assuming randomisation instead is more likely for general circuits and algorithms. 

For the abort datasets shown in \cref{fig:MLE_abort} (orange and red), we emulate $M_\text{retained}$ noisy shots by sampling $M_\text{retained}$ times from the $f$-dependent distribution. For the other datasets, we emulate $M_\text{begun}:=M_\text{retained}/(1-f)$ noisy shots by sampling $M_\text{begun}$ times from the $f=0$ output distribution.

\subsection{Multiplicative Time Cost Caused by QEM}\label{sec:time_overhead}

First, we calculate the multiplicative time cost required to obtain $M_\text{retained}$ shots based on the single-window abort rate.
The advantage of aborting is that it saves time by not letting a circuit-to-be-discarded complete. For an $N$-window circuit which aborts at some point throughout its execution, we want to know, on average, how far the execution proceeds before it is aborted. We define $p_{\text{abort}}(n)$ as the probability that the circuit aborts because of the $n^{\text{th}}$ window. This requires that the current window causes the abort, and additionally that none of the previous windows have caused it. Therefore, $p_{\text{abort}}(n) = {\rho(1-\rho)^{n-1}}/{f}$.
The average window at which the abort happens (provided that it happens) is calculated as $\langle n \rangle = \sum_{n=1}^Nnp_{\text{abort}}(n)$.
We divide this by $N$ to obtain the average fraction of circuit executed before abort:
\begin{equation}
    \frac{`<n>}{N}
    =\frac{1-(1-f)(1+N\rho)}{N\rho f}.
\end{equation}
Let $M_\text{retained}$ be the desired number of shots for the quantum task, and $M_\text{begun}$ the number of attempted shots when the single-window abort rate is $\rho$.
They are related by $(1 -f)M_\text{begun} =M_\text{retained}$.
Of these $M_\text{begun}$ executions, a fraction $1-f$ complete, thus contributing fully to the time duration,
while the other fraction $f$ contributes an average of $\frac{\langle n \rangle}{N}$.
Overall, the mean multiplicative time cost caused by aborting at rate $\rho$ becomes:
\begin{equation}\label{eq:time_overhead}
\omega:=\frac{1-f+f{\langle n \rangle}/{N}}{1-f}=\frac{f/N}{(1-f)[1-(1-f)^{{1}/{N}}]}.
\end{equation}

\Cref{eq:time_overhead} applies to the orange and red datasets in \cref{fig:MLE_abort}, where the setup is that $M_\text{begun}$ shots begin to execute and only $M_\text{retained}$ are allowed to complete, thus saving runtime. However, starting from a minimum number of shots $M_\text{retained}$, a different strategy is to fully execute a larger number of shots $M_\text{begun}$ and not discard any of them, as reflected by the blue, grey, purple, and green datasets in \cref{fig:MLE_abort}. In this case, the multiplicative time factor is simply given by $\omega:=M_\text{begun}/M_\text{retained}:=\frac{1}{1-f}$.

\subsection{Modelling the Distance-19 Surface Code}\label{sec:d_19_model}
In \cref{fig:d=19model} we indicate the expected performance of the methods described in this paper in the context of a $d=19$ surface code. For codes this large, it was not possible to obtain the log success odds $\lambda$ from direct simulation as the computation would be far too costly.
However, we do have available the DCS data shown in \cref{fig:calibration_curve}a.

Therefore, we take the following approach.
For a given decoding window,
we assume the DCS value is a random variable
normally distributed about $\lambda$ with some standard deviation $\sigma_r$,
i.e.~\cref{eq:dcs_as_sum_of_2_rvs}
with $\alpha(\lambda) =\lambda$ and $r \sim \N(0, \sigma_r^2)$.
This is consistent with the behaviour observed in smaller codes;
see Figure S1.

Given this simplifying assumption, we then determine an analytic form for the probability density of $\lambda$ (grey line in \cref{fig:d=19model}), which {\em would indeed give rise to} the observed DCS (green dashed line). Having thus equipped ourselves with analytic forms for both $\lambda$ and the DCS, we can proceed to quickly evaluate multiple scenarios including tasks with billions of decoding windows, as reported in \cref{fig:d=19model}.

Note that by assuming a specific relation between $\lambda$ and the DCS, we are able to perfectly calibrate the DCS: we know exactly the probability distribution over $\lambda$ that is implied by any given observed DCS. Therefore, calibration imperfection does not contribute to the performance reported in \cref{fig:d=19model},
whereas it would in a real system.

The choice of $\sigma_r$ is made as follows. We note that from \cref{fig:source_of_dcs_variation_perturb0.3}, the deviation increases with code size;
however, one cannot say what value it would reach for $d=19$. At $d=11$ it has reached $0.5$ and does not seem to be increasing steeply. In \cref{fig:d=19model} we pessimistically assume that it will reach $\sigma_r=1.0$ for distance $19$, and we also show the improvement that would be achieved
if the DCS were `tighter to the ideal' with $\sigma_r=0.7$.

\subsection{QCELS Parameters}\label{sec:qcels}
Based on numbers reported in \cite{akahoshi_compilation_2024}
for the chosen algorithmic accuracy (the maximum number of Trotter steps involved in any quantum circuit and the average number of clocks per Trotter step), we deduced that the maximum number of decoding windows in any quantum circuit is $\num{ 1.38e9}$.
We use our own numerical predictions for the LER dependence on the code distance at a physical error rate $p_\text{phys}=10^{-3}$, rather than the reference $10^{-4}$. To ensure the noise is sufficiently suppressed, we identify the smallest code distance $d$ such that the mean whole-circuit LEP remains below $10^{-2}$, and that is $d=21$. To obtain a LEP distribution with the same mean as $(N_{\text{windows}}, d) =(\num{ 1.38e9}, 19)$,
we used $(N_{\text{windows}}, d) =(\num{2.57e5}, 11)$.

The maximum allowed single-window LER $r_{\max}$ mentioned under ``Distance-19 Surface Code"
is calculated as follows.
The maximum allowed whole-circuit LER is $R_{\max} =10^{-2}$,
so $r_{\max} =R_{\max}/N_{\text{windows}}\simeq10^{-11}$.

We considered parameters $J=16$ and $N=5$ from Ref.~\cite[Sec.~IV.B]{akahoshi_compilation_2024}
(note the meaning of $N$ here differs from the usual meaning of decoding window count used in this paper).
Therefore, $N_{\text{circuits}}=2JN = 160$ expectation values need to be estimated through different quantum circuits. 
Each circuit should be run $M_\text{retained}$ times, which means that $N_{\text{circuits}}M_\text{retained} = \num{16000}$ quantum circuit runs occur in total.

\subsection{Code Distance Influence on Resources}\label{sec:d_infl_res}

The surface code qubit count scales quadratically with distance $d$,
and the duration of each logical operation scales at most linearly with $d$.
Therefore, the spacetime volume increase is
$(d_\text{small}/d_\text{targ})^3 \omega -1$. For $d_\text{targ}=21$ and $d_\text{small}=19$, under the abort-only protocol, the multiplicative time cost defined under ``Multiplicative Time Cost Caused by QEM" in ``Methods" is $\omega=1.44$, giving a spacetime increase of approximately $7\%$. For MLE-only, $\omega=1.25$, leading to a decrease of approximately $7\%$.

\section*{Data Availability}
All raw data on which the figures and findings of this manuscript are based are publicly available at https://doi.org/10.5281/zenodo.18259317. They are stored in the Zenodo repository titled ``Data for `Error mitigation for logical circuits using decoder confidence','' DOI: 10.5281/zenodo.18259318.

\section*{Code availability}
Custom code and scripts used to generate and analyse the datasets are not publicly available because they have not been deposited in a public repository, but they are available from the corresponding authors upon reasonable request.

\section*{Acknowledgements}
We thank Zi-Han Chen for providing code and guidance on their implementation of the swim distance on MWPM. We also thank Christophe Piveteau for providing code for calculating logical coset probabilities via tensor-network contraction. For useful discussions, we thank Zhenyu Cai, Armands Strikis and Christopher Chubb.

\section*{Funding}
We acknowledge the use of the University of Oxford Advanced Research Computing (ARC) facility~\cite{Richards2015_quantum_bibstyle} in carrying out this work and specifically the facilities made available from the EPSRC QCS Hub grant
(agreement No.\ EP/T001062/1). We also acknowledge support from two EPSRC projects: RoaRQ (EP/W032635/1) and SEEQA (EP/Y004655/1). TC and MD acknowledge support from EPSRC DTP studentships. MD was additionally supported by St Catherine's College, University of Oxford via a Lord Bullock Memorial Scholarship. TC was additionally supported by JST ASPIRE Japan Grant Number JPMJAP2319.

\section*{Author Contributions}
TC conceived and performed the tensor-network comparison reported under ``Decoder Confidence Scores are an Excellent Indicator of Logical Error Risk". MD designed the DCS-MLE procedure and performed the analyses reported under ``Single-Decoding-Window Statistics", ``Many-Decoding-Window Statistics", ``Distance-11 Surface Code", as well as the Hubbard-model resource estimation and the sliding-window analysis in ``Discussion". SCB performed the distance-19 surface-code analysis and supervised the project. All authors contributed equally to writing the manuscript.

\section*{Competing Interests}
The authors declare no competing financial or non-financial interests.

\section*{References}
\bibliographystyle{naturemag}
\bibliography{ref,tchbib,Maria_references}

\begin{thebibliography}{10}
\expandafter\ifx\csname url\endcsname\relax
  \def\url#1{\texttt{#1}}\fi
\expandafter\ifx\csname urlprefix\endcsname\relax\def\urlprefix{URL }\fi
\providecommand{\bibinfo}[2]{#2}
\providecommand{\eprint}[2][]{\url{#2}}

\bibitem{shor_scheme_1995}
\bibinfo{author}{Shor, P.~W.}
\newblock \bibinfo{title}{Scheme for reducing decoherence in quantum computer memory}.
\newblock \emph{\bibinfo{journal}{Physical Review A}} \textbf{\bibinfo{volume}{52}}, \bibinfo{pages}{R2493--R2496} (\bibinfo{year}{1995}).
\newblock \urlprefix\url{https://link.aps.org/doi/10.1103/PhysRevA.52.R2493}.

\bibitem{fowler_surface_2012}
\bibinfo{author}{Fowler, A.~G.}, \bibinfo{author}{Mariantoni, M.}, \bibinfo{author}{Martinis, J.~M.} \& \bibinfo{author}{Cleland, A.~N.}
\newblock \bibinfo{title}{Surface codes: {Towards} practical large-scale quantum computation}.
\newblock \emph{\bibinfo{journal}{Physical Review A}} \textbf{\bibinfo{volume}{86}}, \bibinfo{pages}{032324} (\bibinfo{year}{2012}).
\newblock \urlprefix\url{https://link.aps.org/doi/10.1103/PhysRevA.86.032324}.

\bibitem{Chen2026}
\bibinfo{author}{Chen, Z.-H.}, \bibinfo{author}{Chen, M.-C.}, \bibinfo{author}{Lu, C.-Y.} \& \bibinfo{author}{Pan, J.-W.}
\newblock \bibinfo{title}{Efficient magic state cultivation on $\mathbb{RP}^2$}.
\newblock \emph{\bibinfo{journal}{PRX Quantum}} \textbf{\bibinfo{volume}{7}}, \bibinfo{pages}{010315} (\bibinfo{year}{2026}).

\bibitem{meister_efficient_2024}
\bibinfo{author}{Meister, N.}, \bibinfo{author}{Pattison, C.~A.} \& \bibinfo{author}{Preskill, J.}
\newblock \bibinfo{title}{Efficient soft-output decoders for the surface code} (\bibinfo{year}{2024}).
\newblock \urlprefix\url{http://arxiv.org/abs/2405.07433}.
\newblock \bibinfo{note}{ArXiv:2405.07433 [quant-ph]}.

\bibitem{Lee2025c}
\bibinfo{author}{Lee, S.-H.}, \bibinfo{author}{English, L.} \& \bibinfo{author}{Bartlett, S.~D.}
\newblock \bibinfo{title}{Efficient post-selection for general quantum {LDPC} codes} (\bibinfo{year}{2025}).
\newblock \eprint{2510.05795}.

\bibitem{bombin_fault-tolerant_2024}
\bibinfo{author}{Bombín, H.}, \bibinfo{author}{Pant, M.}, \bibinfo{author}{Roberts, S.} \& \bibinfo{author}{Seetharam, K.~I.}
\newblock \bibinfo{title}{Fault-{Tolerant} {Postselection} for {Low}-{Overhead} {Magic} {State} {Preparation}}.
\newblock \emph{\bibinfo{journal}{PRX Quantum}} \textbf{\bibinfo{volume}{5}}, \bibinfo{pages}{010302} (\bibinfo{year}{2024}).
\newblock \urlprefix\url{https://link.aps.org/doi/10.1103/PRXQuantum.5.010302}.

\bibitem{deMartiiOlius2024}
\bibinfo{author}{deMarti iOlius, A.}, \bibinfo{author}{Fuentes, P.}, \bibinfo{author}{Or{\'{u}}s, R.}, \bibinfo{author}{Crespo, P.~M.} \& \bibinfo{author}{Etxezarreta~Martinez, J.}
\newblock \bibinfo{title}{Decoding algorithms for surface codes}.
\newblock \emph{\bibinfo{journal}{{Quantum}}} \textbf{\bibinfo{volume}{8}}, \bibinfo{pages}{1498} (\bibinfo{year}{2024}).

\bibitem{Gidney2025}
\bibinfo{author}{Gidney, C.}, \bibinfo{author}{Newman, M.}, \bibinfo{author}{Brooks, P.} \& \bibinfo{author}{Jones, C.}
\newblock \bibinfo{title}{Yoked surface codes}.
\newblock \emph{\bibinfo{journal}{Nature Communications}} \textbf{\bibinfo{volume}{16}}, \bibinfo{pages}{4498} (\bibinfo{year}{2025}).

\bibitem{Pattison2025}
\bibinfo{author}{Pattison, C.~A.}, \bibinfo{author}{Krishna, A.} \& \bibinfo{author}{Preskill, J.}
\newblock \bibinfo{title}{Hierarchical memories: Simulating quantum {LDPC} codes with local gates}.
\newblock \emph{\bibinfo{journal}{{Quantum}}} \textbf{\bibinfo{volume}{9}}, \bibinfo{pages}{1728} (\bibinfo{year}{2025}).

\bibitem{Akahoshi2025}
\bibinfo{author}{Akahoshi, Y.} \emph{et~al.}
\newblock \bibinfo{title}{Runtime reduction in lattice surgery utilizing time-like soft information} (\bibinfo{year}{2025}).
\newblock \eprint{2510.21149}.

\bibitem{Sunami2025}
\bibinfo{author}{Sunami, S.}, \bibinfo{author}{Hirano, Y.}, \bibinfo{author}{Hinokuma, T.} \& \bibinfo{author}{Yamasaki, H.}
\newblock \bibinfo{title}{Entanglement boosting: Low-volume logical {Bell} pair preparation for distributed fault-tolerant quantum computation} (\bibinfo{year}{2025}).
\newblock \eprint{2511.10729}.

\bibitem{Poulin2006}
\bibinfo{author}{Poulin, D.}
\newblock \bibinfo{title}{Optimal and efficient decoding of concatenated quantum block codes}.
\newblock \emph{\bibinfo{journal}{Physical Review A}} \textbf{\bibinfo{volume}{74}}, \bibinfo{pages}{052333} (\bibinfo{year}{2006}).

\bibitem{Goto2013}
\bibinfo{author}{Goto, H.} \& \bibinfo{author}{Uchikawa, H.}
\newblock \bibinfo{title}{Fault-tolerant quantum computation with a soft-decision decoder for error correction and detection by teleportation}.
\newblock \emph{\bibinfo{journal}{Scientific Reports}} \textbf{\bibinfo{volume}{3}}, \bibinfo{pages}{2044} (\bibinfo{year}{2013}).

\bibitem{siegel_snakes_2025}
\bibinfo{author}{Siegel, A.} \emph{et~al.}
\newblock \bibinfo{title}{Snakes on a {Plane}: mobile, low dimensional logical qubits on a {2D} surface} (\bibinfo{year}{2025}).
\newblock \urlprefix\url{http://arxiv.org/abs/2501.02120}.
\newblock \bibinfo{note}{ArXiv:2501.02120 [quant-ph]}.

\bibitem{Smith2024}
\bibinfo{author}{Smith, S.~C.}, \bibinfo{author}{Brown, B.~J.} \& \bibinfo{author}{Bartlett, S.~D.}
\newblock \bibinfo{title}{Mitigating errors in logical qubits}.
\newblock \emph{\bibinfo{journal}{Communications Physics}} \textbf{\bibinfo{volume}{7}}, \bibinfo{pages}{386} (\bibinfo{year}{2024}).

\bibitem{SalesRodriguez2025}
\bibinfo{author}{Sales~Rodriguez, P.} \emph{et~al.}
\newblock \bibinfo{title}{Experimental demonstration of logical magic state distillation}.
\newblock \emph{\bibinfo{journal}{Nature}} \textbf{\bibinfo{volume}{645}}, \bibinfo{pages}{620--625} (\bibinfo{year}{2025}).

\bibitem{gidney_magic_2024}
\bibinfo{author}{Gidney, C.}, \bibinfo{author}{Shutty, N.} \& \bibinfo{author}{Jones, C.}
\newblock \bibinfo{title}{Magic state cultivation: growing {T} states as cheap as {CNOT} gates} (\bibinfo{year}{2024}).
\newblock \urlprefix\url{http://arxiv.org/abs/2409.17595}.
\newblock \bibinfo{note}{ArXiv:2409.17595 [quant-ph]}.

\bibitem{Sahay2025}
\bibinfo{author}{Sahay, K.} \emph{et~al.}
\newblock \bibinfo{title}{Fold-transversal surface code cultivation} (\bibinfo{year}{2025}).
\newblock \eprint{2509.05212}.

\bibitem{bluvsteinLogicalQuantumProcessor2024}
\bibinfo{author}{Bluvstein, D.} \emph{et~al.}
\newblock \bibinfo{title}{Logical quantum processor based on reconfigurable atom arrays}.
\newblock \emph{\bibinfo{journal}{Nature}} \textbf{\bibinfo{volume}{626}}, \bibinfo{pages}{58--65} (\bibinfo{year}{2024}).
\newblock \urlprefix\url{https://www.nature.com/articles/s41586-023-06927-3}.

\bibitem{Shutty2024}
\bibinfo{author}{Shutty, N.}, \bibinfo{author}{Newman, M.} \& \bibinfo{author}{Villalonga, B.}
\newblock \bibinfo{title}{Efficient near-optimal decoding of the surface code through ensembling} (\bibinfo{year}{2024}).
\newblock \eprint{2401.12434}.

\bibitem{Toshio2025}
\bibinfo{author}{Toshio, R.} \emph{et~al.}
\newblock \bibinfo{title}{Decoder switching: Breaking the speed-accuracy tradeoff in real-time quantum error correction} (\bibinfo{year}{2025}).
\newblock \eprint{2510.25222}.

\bibitem{Birchall2026}
\bibinfo{author}{Birchall, P.} \emph{et~al.}
\newblock \bibinfo{title}{{Macromux}: scalable postselection for high-threshold fault-tolerant quantum computation} (\bibinfo{year}{2026}).
\newblock \eprint{2603.04875}.

\bibitem{Staples2026}
\bibinfo{author}{Staples, J.~W.}, \bibinfo{author}{Fu, W.} \& \bibinfo{author}{Thompson, J.~D.}
\newblock \bibinfo{title}{Scalable postselection of quantum resources} (\bibinfo{year}{2026}).
\newblock \eprint{2603.08697}.

\bibitem{Oberoi2026}
\bibinfo{author}{Oberoi, T.}, \bibinfo{author}{Viszlai, J.} \& \bibinfo{author}{Chong, F.~T.}
\newblock \bibinfo{title}{{ADaPT}: Adaptive-window decoding for practical fault-tolerance} (\bibinfo{year}{2026}).
\newblock \eprint{2605.01149}.

\bibitem{Mishima2026}
\bibinfo{author}{Mishima, M.} \emph{et~al.}
\newblock \bibinfo{title}{Adaptive window decoding based on spatiotemporal complementary gap} (\bibinfo{year}{2026}).
\newblock \eprint{2605.14637}.

\bibitem{cai_quantum_2023}
\bibinfo{author}{Cai, Z.} \emph{et~al.}
\newblock \bibinfo{title}{Quantum error mitigation}.
\newblock \emph{\bibinfo{journal}{Reviews of Modern Physics}} \textbf{\bibinfo{volume}{95}}, \bibinfo{pages}{045005} (\bibinfo{year}{2023}).
\newblock \urlprefix\url{https://link.aps.org/doi/10.1103/RevModPhys.95.045005}.

\bibitem{liEfficientVariationalQuantum2017}
\bibinfo{author}{Li, Y.} \& \bibinfo{author}{Benjamin, S.~C.}
\newblock \bibinfo{title}{Efficient variational quantum simulator incorporating active error minimization}.
\newblock \emph{\bibinfo{journal}{Phys. Rev. X}} \textbf{\bibinfo{volume}{7}}, \bibinfo{pages}{21050} (\bibinfo{year}{2017}).
\newblock \urlprefix\url{https://link.aps.org/doi/10.1103/PhysRevX.7.021050}.

\bibitem{temmeErrorMitigationShortdepth2017}
\bibinfo{author}{Temme, K.}, \bibinfo{author}{Bravyi, S.} \& \bibinfo{author}{Gambetta, J.~M.}
\newblock \bibinfo{title}{Error mitigation for short-depth quantum circuits}.
\newblock \emph{\bibinfo{journal}{Phys. Rev. Lett.}} \textbf{\bibinfo{volume}{119}}, \bibinfo{pages}{180509} (\bibinfo{year}{2017}).
\newblock \urlprefix\url{https://link.aps.org/doi/10.1103/PhysRevLett.119.180509}.

\bibitem{giurgica-tironDigitalZeroNoise2020}
\bibinfo{author}{{Giurgica-Tiron}, T.}, \bibinfo{author}{Hindy, Y.}, \bibinfo{author}{LaRose, R.}, \bibinfo{author}{Mari, A.} \& \bibinfo{author}{Zeng, W.~J.}
\newblock \bibinfo{title}{Digital zero noise extrapolation for quantum error mitigation}.
\newblock In \emph{\bibinfo{booktitle}{2020 {{IEEE International Conference}} on {{Quantum Computing}} and {{Engineering}} ({{QCE}})}}, \bibinfo{pages}{306--316} (\bibinfo{year}{2020}).

\bibitem{endo_practical_2018}
\bibinfo{author}{Endo, S.}, \bibinfo{author}{Benjamin, S.~C.} \& \bibinfo{author}{Li, Y.}
\newblock \bibinfo{title}{Practical {Quantum} {Error} {Mitigation} for {Near}-{Future} {Applications}}.
\newblock \emph{\bibinfo{journal}{Physical Review X}} \textbf{\bibinfo{volume}{8}}, \bibinfo{pages}{031027} (\bibinfo{year}{2018}).
\newblock \urlprefix\url{https://link.aps.org/doi/10.1103/PhysRevX.8.031027}.

\bibitem{mariExtendingQuantumProbabilistic2021}
\bibinfo{author}{Mari, A.}, \bibinfo{author}{Shammah, N.} \& \bibinfo{author}{Zeng, W.~J.}
\newblock \bibinfo{title}{Extending quantum probabilistic error cancellation by noise scaling}.
\newblock \emph{\bibinfo{journal}{Phys. Rev. A}} \textbf{\bibinfo{volume}{104}}, \bibinfo{pages}{52607} (\bibinfo{year}{2021}).
\newblock \urlprefix\url{https://link.aps.org/doi/10.1103/PhysRevA.104.052607}.

\bibitem{moske_random-matrix_2026}
\bibinfo{author}{Moske, L.} \emph{et~al.}
\newblock \bibinfo{title}{Random-matrix perspective on probabilistic error cancellation}.
\newblock \emph{\bibinfo{journal}{Physical Review A}} \textbf{\bibinfo{volume}{114}}, \bibinfo{pages}{012442} (\bibinfo{year}{2026}).
\newblock \urlprefix\url{https://link.aps.org/doi/10.1103/5q85-fp2p}.

\bibitem{bonet-monroigLowcostErrorMitigation2018}
\bibinfo{author}{{Bonet-Monroig}, X.}, \bibinfo{author}{Sagastizabal, R.}, \bibinfo{author}{Singh, M.} \& \bibinfo{author}{O'Brien, T.~E.}
\newblock \bibinfo{title}{Low-cost error mitigation by symmetry verification}.
\newblock \emph{\bibinfo{journal}{Phys. Rev. A}} \textbf{\bibinfo{volume}{98}}, \bibinfo{pages}{62339} (\bibinfo{year}{2018}).
\newblock \urlprefix\url{https://link.aps.org/doi/10.1103/PhysRevA.98.062339}.

\bibitem{mcardleErrormitigatedDigitalQuantum2019}
\bibinfo{author}{McArdle, S.}, \bibinfo{author}{Yuan, X.} \& \bibinfo{author}{Benjamin, S.}
\newblock \bibinfo{title}{Error-mitigated digital quantum simulation}.
\newblock \emph{\bibinfo{journal}{Phys. Rev. Lett.}} \textbf{\bibinfo{volume}{122}}, \bibinfo{pages}{180501} (\bibinfo{year}{2019}).
\newblock \urlprefix\url{https://link.aps.org/doi/10.1103/PhysRevLett.122.180501}.

\bibitem{caiQuantumErrorMitigation2021}
\bibinfo{author}{Cai, Z.}
\newblock \bibinfo{title}{Quantum error mitigation using symmetry expansion}.
\newblock \emph{\bibinfo{journal}{Quantum}} \textbf{\bibinfo{volume}{5}}, \bibinfo{pages}{548} (\bibinfo{year}{2021}).
\newblock \urlprefix\url{https://quantum-journal.org/papers/q-2021-09-21-548/}.

\bibitem{hugginsVirtualDistillationQuantum2021}
\bibinfo{author}{Huggins, W.~J.} \emph{et~al.}
\newblock \bibinfo{title}{Virtual distillation for quantum error mitigation}.
\newblock \emph{\bibinfo{journal}{Phys. Rev. X}} \textbf{\bibinfo{volume}{11}}, \bibinfo{pages}{41036} (\bibinfo{year}{2021}).
\newblock \urlprefix\url{https://link.aps.org/doi/10.1103/PhysRevX.11.041036}.

\bibitem{koczorExponentialErrorSuppression2021}
\bibinfo{author}{Koczor, B.}
\newblock \bibinfo{title}{Exponential error suppression for near-term quantum devices}.
\newblock \emph{\bibinfo{journal}{Phys. Rev. X}} \textbf{\bibinfo{volume}{11}}, \bibinfo{pages}{31057} (\bibinfo{year}{2021}).
\newblock \urlprefix\url{https://link.aps.org/doi/10.1103/PhysRevX.11.031057}.

\bibitem{obrienPurificationbasedQuantumError2023}
\bibinfo{author}{O'Brien, T.~E.} \emph{et~al.}
\newblock \bibinfo{title}{Purification-based quantum error mitigation of pair-correlated electron simulations}.
\newblock \emph{\bibinfo{journal}{Nat. Phys.}} \bibinfo{pages}{1--6} (\bibinfo{year}{2023}).
\newblock \urlprefix\url{https://www.nature.com/articles/s41567-023-02240-y}.

\bibitem{czarnikErrorMitigationClifford2021}
\bibinfo{author}{Czarnik, P.}, \bibinfo{author}{Arrasmith, A.}, \bibinfo{author}{Coles, P.~J.} \& \bibinfo{author}{Cincio, L.}
\newblock \bibinfo{title}{Error mitigation with clifford quantum-circuit data}.
\newblock \emph{\bibinfo{journal}{Quantum}} \textbf{\bibinfo{volume}{5}}, \bibinfo{pages}{592} (\bibinfo{year}{2021}).
\newblock \urlprefix\url{https://quantum-journal.org/papers/q-2021-11-26-592/}.

\bibitem{strikisLearningbasedQuantumError2021}
\bibinfo{author}{Strikis, A.}, \bibinfo{author}{Qin, D.}, \bibinfo{author}{Chen, Y.}, \bibinfo{author}{Benjamin, S.~C.} \& \bibinfo{author}{Li, Y.}
\newblock \bibinfo{title}{Learning-based quantum error mitigation}.
\newblock \emph{\bibinfo{journal}{PRX Quantum}} \textbf{\bibinfo{volume}{2}}, \bibinfo{pages}{40330} (\bibinfo{year}{2021}).
\newblock \urlprefix\url{https://link.aps.org/doi/10.1103/PRXQuantum.2.040330}.

\bibitem{vandenbergProbabilisticErrorCancellation2023}
\bibinfo{author}{{van den Berg}, E.}, \bibinfo{author}{Minev, Z.~K.}, \bibinfo{author}{Kandala, A.} \& \bibinfo{author}{Temme, K.}
\newblock \bibinfo{title}{Probabilistic error cancellation with sparse pauli--lindblad models on noisy quantum processors}.
\newblock \emph{\bibinfo{journal}{Nat. Phys.}} \bibinfo{pages}{1--6} (\bibinfo{year}{2023}).
\newblock \urlprefix\url{https://www.nature.com/articles/s41567-023-02042-2}.

\bibitem{kim_evidence_2023}
\bibinfo{author}{Kim, Y.} \emph{et~al.}
\newblock \bibinfo{title}{Evidence for the utility of quantum computing before fault tolerance}.
\newblock \emph{\bibinfo{journal}{Nature}} \textbf{\bibinfo{volume}{618}}, \bibinfo{pages}{500--505} (\bibinfo{year}{2023}).
\newblock \urlprefix\url{https://www.nature.com/articles/s41586-023-06096-3}.

\bibitem{leviatan_resolving_2026}
\bibinfo{author}{Leviatan, E.} \emph{et~al.}
\newblock \bibinfo{title}{Resolving {Structure} in {Prethermal} {Floquet} {Dynamics} with {Precision} {Quantum} {Computation}} (\bibinfo{year}{2026}).
\newblock \urlprefix\url{http://arxiv.org/abs/2607.24937}.
\newblock \bibinfo{note}{ArXiv:2607.24937 [quant-ph]}.

\bibitem{suzuki_quantum_2022}
\bibinfo{author}{Suzuki, Y.}, \bibinfo{author}{Endo, S.}, \bibinfo{author}{Fujii, K.} \& \bibinfo{author}{Tokunaga, Y.}
\newblock \bibinfo{title}{Quantum {Error} {Mitigation} as a {Universal} {Error} {Reduction} {Technique}: {Applications} from the {NISQ} to the {Fault}-{Tolerant} {Quantum} {Computing} {Eras}}.
\newblock \emph{\bibinfo{journal}{PRX Quantum}} \textbf{\bibinfo{volume}{3}}, \bibinfo{pages}{010345} (\bibinfo{year}{2022}).
\newblock \urlprefix\url{https://link.aps.org/doi/10.1103/PRXQuantum.3.010345}.

\bibitem{aharonov_importance_2025}
\bibinfo{author}{Aharonov, D.} \emph{et~al.}
\newblock \bibinfo{title}{On the {Importance} of {Error} {Mitigation} for {Quantum} {Computation}} (\bibinfo{year}{2025}).
\newblock \urlprefix\url{http://arxiv.org/abs/2503.17243}.
\newblock \bibinfo{note}{ArXiv:2503.17243 [quant-ph]}.

\bibitem{zhang_demonstrating_2025}
\bibinfo{author}{Zhang, A.} \emph{et~al.}
\newblock \bibinfo{title}{Demonstrating quantum error mitigation on logical qubits} (\bibinfo{year}{2025}).
\newblock \urlprefix\url{http://arxiv.org/abs/2501.09079}.
\newblock \bibinfo{note}{ArXiv:2501.09079 [quant-ph]}.

\bibitem{jeon_quantum_2026}
\bibinfo{author}{Jeon, M.} \& \bibinfo{author}{Cai, Z.}
\newblock \bibinfo{title}{Quantum {Error} {Correction} on {Error}-mitigated {Physical} {Qubits}} (\bibinfo{year}{2026}).
\newblock \urlprefix\url{http://arxiv.org/abs/2601.18384}.
\newblock \bibinfo{note}{ArXiv:2601.18384 [quant-ph]}.

\bibitem{dutkiewicz_error_2024}
\bibinfo{author}{Dutkiewicz, A.} \emph{et~al.}
\newblock \bibinfo{title}{Error mitigation and circuit division for early fault-tolerant quantum phase estimation} (\bibinfo{year}{2024}).
\newblock \urlprefix\url{http://arxiv.org/abs/2410.05369}.
\newblock \bibinfo{note}{ArXiv:2410.05369 [quant-ph]}.

\bibitem{wahl_zero_2023}
\bibinfo{author}{Wahl, M.~A.}, \bibinfo{author}{Mari, A.}, \bibinfo{author}{Shammah, N.}, \bibinfo{author}{Zeng, W.~J.} \& \bibinfo{author}{Ravi, G.~S.}
\newblock \bibinfo{title}{Zero {Noise} {Extrapolation} on {Logical} {Qubits} by {Scaling} the {Error} {Correction} {Code} {Distance}}.
\newblock In \emph{\bibinfo{booktitle}{2023 {IEEE} {International} {Conference} on {Quantum} {Computing} and {Engineering} ({QCE})}}, \bibinfo{pages}{888--897} (\bibinfo{publisher}{IEEE}, \bibinfo{address}{Bellevue, WA, USA}, \bibinfo{year}{2023}).
\newblock \urlprefix\url{https://ieeexplore.ieee.org/document/10313873/}.

\bibitem{umbrarescu_infinite_2026}
\bibinfo{author}{Umbrarescu, G.}, \bibinfo{author}{Higgott, O.} \& \bibinfo{author}{Browne, D.~E.}
\newblock \bibinfo{title}{Infinite {Distance} {Extrapolation}: {How} error mitigation can enhance quantum error correction} (\bibinfo{year}{2026}).
\newblock \urlprefix\url{http://arxiv.org/abs/2603.11285}.
\newblock \bibinfo{note}{ArXiv:2603.11285 [quant-ph] version: 1}.

\bibitem{Dennis2002}
\bibinfo{author}{Dennis, E.}, \bibinfo{author}{Kitaev, A.}, \bibinfo{author}{Landahl, A.} \& \bibinfo{author}{Preskill, J.}
\newblock \bibinfo{title}{Topological quantum memory}.
\newblock \emph{\bibinfo{journal}{Journal of Mathematical Physics}} \textbf{\bibinfo{volume}{43}}, \bibinfo{pages}{4452--4505} (\bibinfo{year}{2002}).

\bibitem{Hutter2014}
\bibinfo{author}{Hutter, A.}, \bibinfo{author}{Wootton, J.~R.} \& \bibinfo{author}{Loss, D.}
\newblock \bibinfo{title}{Efficient {Markov} chain {Monte} {Carlo} algorithm for the surface code}.
\newblock \emph{\bibinfo{journal}{Physical Review A}} \textbf{\bibinfo{volume}{89}}, \bibinfo{pages}{022326} (\bibinfo{year}{2014}).

\bibitem{piveteau_tensor-network_2024}
\bibinfo{author}{Piveteau, C.}, \bibinfo{author}{Chubb, C.~T.} \& \bibinfo{author}{Renes, J.~M.}
\newblock \bibinfo{title}{Tensor-{Network} {Decoding} {Beyond} {2D}}.
\newblock \emph{\bibinfo{journal}{PRX Quantum}} \textbf{\bibinfo{volume}{5}}, \bibinfo{pages}{040303} (\bibinfo{year}{2024}).
\newblock \urlprefix\url{https://link.aps.org/doi/10.1103/PRXQuantum.5.040303}.

\bibitem{Chen2025a}
\bibinfo{author}{Chen, H.} \emph{et~al.}
\newblock \bibinfo{title}{Scalable accuracy gains from postselection in quantum error correcting codes} (\bibinfo{year}{2025}).
\newblock \eprint{2510.05222}.

\bibitem{English2025}
\bibinfo{author}{English, L.~H.}, \bibinfo{author}{Williamson, D.~J.} \& \bibinfo{author}{Bartlett, S.~D.}
\newblock \bibinfo{title}{Thresholds for postselected quantum error correction from statistical mechanics}.
\newblock \emph{\bibinfo{journal}{Physical Review Letters}} \textbf{\bibinfo{volume}{135}}, \bibinfo{pages}{120603} (\bibinfo{year}{2025}).

\bibitem{Higgott2025}
\bibinfo{author}{Higgott, O.} \& \bibinfo{author}{Gidney, C.}
\newblock \bibinfo{title}{{Sparse} {Blossom}: correcting a million errors per core second with minimum-weight matching}.
\newblock \emph{\bibinfo{journal}{{Quantum}}} \textbf{\bibinfo{volume}{9}}, \bibinfo{pages}{1600} (\bibinfo{year}{2025}).

\bibitem{Gidney2021a}
\bibinfo{author}{Gidney, C.}
\newblock \bibinfo{title}{{Stim}: a fast stabilizer circuit simulator}.
\newblock \emph{\bibinfo{journal}{{Quantum}}} \textbf{\bibinfo{volume}{5}}, \bibinfo{pages}{497} (\bibinfo{year}{2021}).

\bibitem{Knill2005}
\bibinfo{author}{Knill, E.}
\newblock \bibinfo{title}{Quantum computing with realistically noisy devices}.
\newblock \emph{\bibinfo{journal}{Nature}} \textbf{\bibinfo{volume}{434}}, \bibinfo{pages}{39--44} (\bibinfo{year}{2005}).

\bibitem{helstrom_quantum_1969}
\bibinfo{author}{Helstrom, C.~W.}
\newblock \bibinfo{title}{Quantum detection and estimation theory}.
\newblock \emph{\bibinfo{journal}{Journal of Statistical Physics}} \textbf{\bibinfo{volume}{1}}, \bibinfo{pages}{231--252} (\bibinfo{year}{1969}).
\newblock \urlprefix\url{https://doi.org/10.1007/BF01007479}.

\bibitem{holevo_probabilistic_2011}
\bibinfo{author}{Holevo, A.}
\newblock \emph{\bibinfo{title}{Probabilistic and {Statistical} {Aspects} of {Quantum} {Theory}}} (\bibinfo{publisher}{Edizioni della Normale}, \bibinfo{address}{Pisa}, \bibinfo{year}{2011}).
\newblock \urlprefix\url{http://link.springer.com/10.1007/978-88-7642-378-9}.

\bibitem{tsubouchi_quantum_2026}
\bibinfo{author}{Tsubouchi, K.}, \bibinfo{author}{Kwon, H.}, \bibinfo{author}{Jiang, L.} \& \bibinfo{author}{Yoshioka, N.}
\newblock \bibinfo{title}{Quantum advantages for syndrome-aware noisy logical observable estimation} (\bibinfo{year}{2026}).
\newblock \urlprefix\url{http://arxiv.org/abs/2603.05145}.
\newblock \bibinfo{note}{ArXiv:2603.05145 [quant-ph] version: 2}.

\bibitem{ding_even_2023}
\bibinfo{author}{Ding, Z.} \& \bibinfo{author}{Lin, L.}
\newblock \bibinfo{title}{Even {Shorter} {Quantum} {Circuit} for {Phase} {Estimation} on {Early} {Fault}-{Tolerant} {Quantum} {Computers} with {Applications} to {Ground}-{State} {Energy} {Estimation}}.
\newblock \emph{\bibinfo{journal}{PRX Quantum}} \textbf{\bibinfo{volume}{4}}, \bibinfo{pages}{020331} (\bibinfo{year}{2023}).
\newblock \urlprefix\url{https://link.aps.org/doi/10.1103/PRXQuantum.4.020331}.

\bibitem{akahoshi_compilation_2024}
\bibinfo{author}{Akahoshi, Y.} \emph{et~al.}
\newblock \bibinfo{title}{Compilation of {Trotter}-{Based} {Time} {Evolution} for {Partially} {Fault}-{Tolerant} {Quantum} {Computing} {Architecture}} (\bibinfo{year}{2024}).
\newblock \urlprefix\url{http://arxiv.org/abs/2408.14929}.
\newblock \bibinfo{note}{ArXiv:2408.14929 [quant-ph]}.

\bibitem{Zhou2025b}
\bibinfo{author}{Zhou, Z.}, \bibinfo{author}{Pexton, S.}, \bibinfo{author}{Kubica, A.} \& \bibinfo{author}{Ding, Y.}
\newblock \bibinfo{title}{Error mitigation of fault-tolerant quantum circuits with soft information} (\bibinfo{year}{2025}).
\newblock \eprint{2512.09863}.

\bibitem{aharonov_syndrome_2025}
\bibinfo{author}{Aharonov, D.} \emph{et~al.}
\newblock \bibinfo{title}{Syndrome aware mitigation of logical errors} (\bibinfo{year}{2025}).
\newblock \urlprefix\url{http://arxiv.org/abs/2512.23810}.
\newblock \bibinfo{note}{ArXiv:2512.23810 [quant-ph]}.

\bibitem{Skoric2023}
\bibinfo{author}{Skoric, L.}, \bibinfo{author}{Browne, D.~E.}, \bibinfo{author}{Barnes, K.~M.}, \bibinfo{author}{Gillespie, N.~I.} \& \bibinfo{author}{Campbell, E.~T.}
\newblock \bibinfo{title}{Parallel window decoding enables scalable fault tolerant quantum computation}.
\newblock \emph{\bibinfo{journal}{Nature Communications}} \textbf{\bibinfo{volume}{14}}, \bibinfo{pages}{7040} (\bibinfo{year}{2023}).

\bibitem{Bombin2023}
\bibinfo{author}{Bomb{\'\i}n, H.} \emph{et~al.}
\newblock \bibinfo{title}{Modular decoding: parallelizable real-time decoding for quantum computers} (\bibinfo{year}{2023}).
\newblock \eprint{2303.04846}.

\bibitem{Chan2026}
\bibinfo{author}{Chan, T.}
\newblock \bibinfo{title}{{Snowflake}: A distributed streaming decoder}.
\newblock \emph{\bibinfo{journal}{{Quantum}}} \textbf{\bibinfo{volume}{10}}, \bibinfo{pages}{2033} (\bibinfo{year}{2026}).

\bibitem{Lin2025a}
\bibinfo{author}{Fuhui~Lin, S.}, \bibinfo{author}{Peterson, E.~C.}, \bibinfo{author}{Sankar, K.} \& \bibinfo{author}{Sivarajah, P.}
\newblock \bibinfo{title}{Spatially parallel decoding for multi-qubit lattice surgery}.
\newblock \emph{\bibinfo{journal}{Quantum Science and Technology}} \textbf{\bibinfo{volume}{10}}, \bibinfo{pages}{035007} (\bibinfo{year}{2025}).

\bibitem{gu_scalable_2026}
\bibinfo{author}{Gu, A.}, \bibinfo{author}{Ataides, J. P.~B.}, \bibinfo{author}{Lukin, M.~D.} \& \bibinfo{author}{Yelin, S.~F.}
\newblock \bibinfo{title}{Scalable {Neural} {Decoders} for {Practical} {Fault}-{Tolerant} {Quantum} {Computation}} (\bibinfo{year}{2026}).
\newblock \urlprefix\url{http://arxiv.org/abs/2604.08358}.
\newblock \bibinfo{note}{ArXiv:2604.08358 [quant-ph]}.

\bibitem{Dentelski2026}
\bibinfo{author}{Dentelski, D.}
\newblock \bibinfo{title}{Neural network decoder confidence as a learned proxy for the logical gap} (\bibinfo{year}{2026}).
\newblock \eprint{2606.08758}.

\bibitem{wills_forced_2026}
\bibinfo{author}{Wills, A.}, \bibinfo{author}{Yoder, T.~J.} \& \bibinfo{author}{Chuang, I.}
\newblock \bibinfo{title}{Forced {Gap} {Post}-{Selection} for {Quantum} {LDPC} {Codes} and their {Operations}} (\bibinfo{year}{2026}).
\newblock \urlprefix\url{http://arxiv.org/abs/2605.20346}.
\newblock \bibinfo{note}{ArXiv:2605.20346 [quant-ph]}.

\bibitem{Richards2015_quantum_bibstyle}
\bibinfo{author}{Richards, A.}
\newblock \emph{\bibinfo{title}{\href{https://doi.org/10.5281/zenodo.22558}{University of Oxford Advanced Research Computing}}} (\bibinfo{year}{2015}).

\end{thebibliography}

\clearpage
\onecolumngrid
\setcounter{figure}{0}
\setcounter{equation}{0}
\renewcommand{\thefigure}{S\arabic{figure}}
\renewcommand{\theequation}{S\arabic{equation}}

\section*{Supplementary Information}
\section*{Supplementary Note 1: Additional diagnostic figures}
The following figures provide the calibration and convergence diagnostics referenced from the Methods of the main manuscript.

\begin{figure}[ht]
\centering
\includegraphics[width=0.82\linewidth]{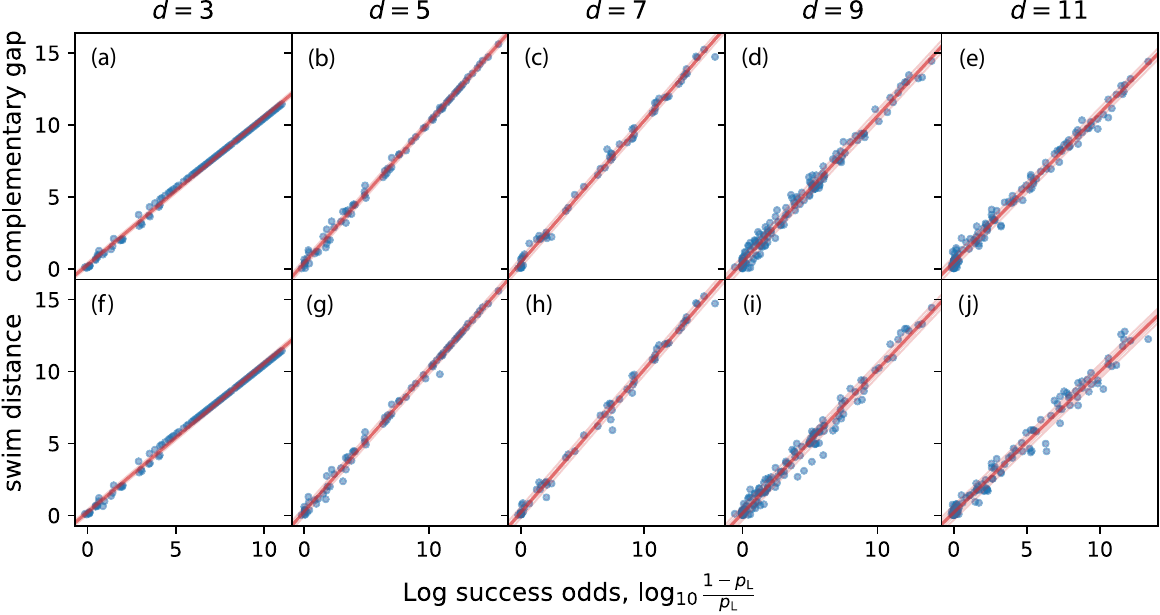}
\caption{\textbf{Decoder confidence scores are compared with the log success odds for different code distances.} Panels (a)--(e) show the complementary gap for $d=3,5,7,9,11$, respectively, and panels (f)--(j) show the swim distance for the same code distances. Each blue point is one syndrome sample from the unrotated surface code under the perturbed phenomenological noise model, decoded using MWPM. The red line is the unweighted least-squares best fit; the shaded region has height equal to the root mean squared residual $s_r$. Due to floating-point precision, $\lambda$ is capped at 16, so for $d\ge9$ samples whose complementary gap exceeds 16 are omitted.}
\label{fig:s1}
\end{figure}

\begin{figure}[ht]
\centering
\includegraphics[width=0.72\linewidth]{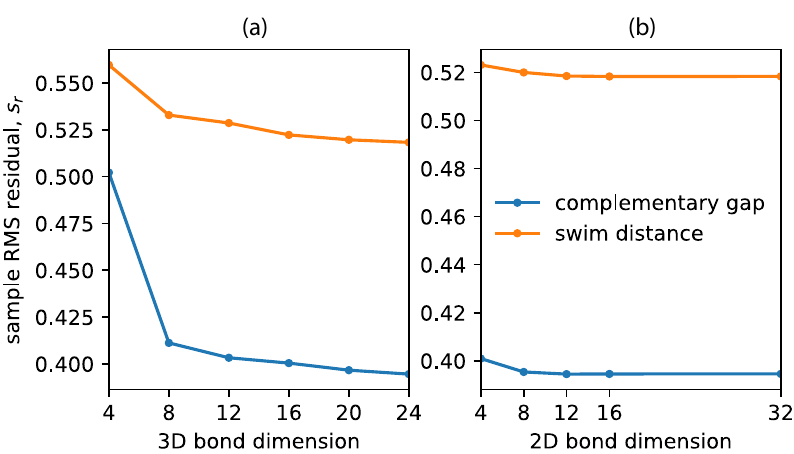}
\caption{\textbf{Tensor-network contraction convergence is assessed through the sample root mean squared residual.} (a) The residual $s_r$ is plotted against the bond dimension used in the 3D-to-2D contraction stage. (b) The residual is plotted against the bond dimension used in the subsequent 2D contraction stage. Results are for the distance-11 surface code; blue and orange curves correspond to the complementary gap and swim distance, respectively.}
\label{fig:s2}
\end{figure}

\begin{figure}[ht]
\centering
\includegraphics[width=0.62\linewidth]{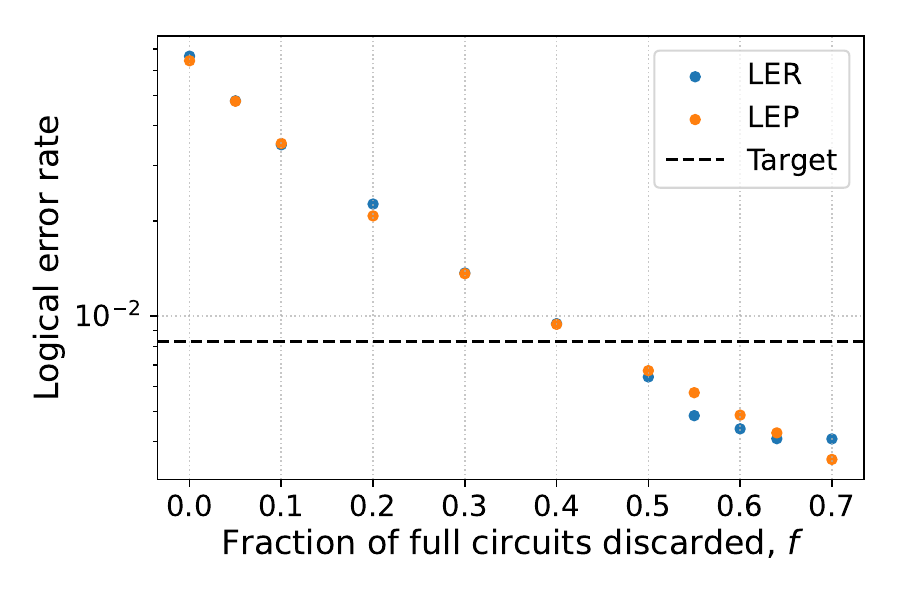}
\caption{\textbf{Logical error rate and mean logical error probability decrease as more full-circuit executions are discarded.} Results are for $d=11$, obtained by sampling $N=2.57\times10^5$ single-window $(x,p_\mathrm{L})$ pairs $5\times10^6$ times. The solid curve is the mean logical error probability, the dashed curve is the observed logical error rate, and the horizontal target line is specific to the Hubbard-model application in the main manuscript.}
\label{fig:s3}
\end{figure}

\clearpage
\section*{Supplementary Note 2: LEP Variance of a Many-Window Experiment}
As discussed in the main text,
the total LEP for an experiment comprising $N$ windows,
each with LEP $p_{\L, i}$ is given by Eq. (3) of the main manuscript.
If all $p_{\L,i}$ have mean $\mu_{1,d}$ and variance $\sigma_{1,d}^2$,
the random variable $P_\L$ has mean and variance:
\begin{align}
    \mu_{N,d} &= \tfrac12\big[1 - (1 -2\mu_{1,d})^N\big], \\
    \sigma_{N,d}^2 &= \tfrac14\big\{\big[(1 -2\mu_{1,d})^{2} +(2\sigma_{1,d})^2\big]^N - (1 -2\mu_{1,d})^{2N}\big\}.
\end{align}

Figure S4a shows that
both $\mu_{1,d}$ and $\sigma_{1,d}^2$ are very small;
in this case,
$\sigma_{N,d}^2$ vanishes at constant $d$ as $N\rightarrow\infty$.
In practical contexts, however, the distribution of $P_\L$ remains broad (and thus useful) as circuit sizes increase. The explanation lies in the observation that, as $N$ increases, the mean total LEP $\mu_{N,d}$ also increases at constant $d$, possibly exceeding the maximum error allowed for the particular application under consideration;
see Figure S4b.
The solution is to increase the code size: for a fixed target LER $\epsilon$, we choose the smallest code distance $d(\epsilon)$ that fulfils $\mu_{N,d(\epsilon)} \leq \epsilon$. The question then becomes: how does $\sigma_{N,d(\epsilon)}^2$ behave?
Figure S4c indicates that this variance does not shrink as $N$ increases. In this subsection, we replaced the previously used $\bar{p}_\L$ and $\bar{P}_\L$ variables by $\mu_{1,d}$ and $\mu_{N,d}$, to highlight their dependence on the code distance and window number.

\begin{figure}[ht]
\centering
\includegraphics[width=0.98\linewidth]{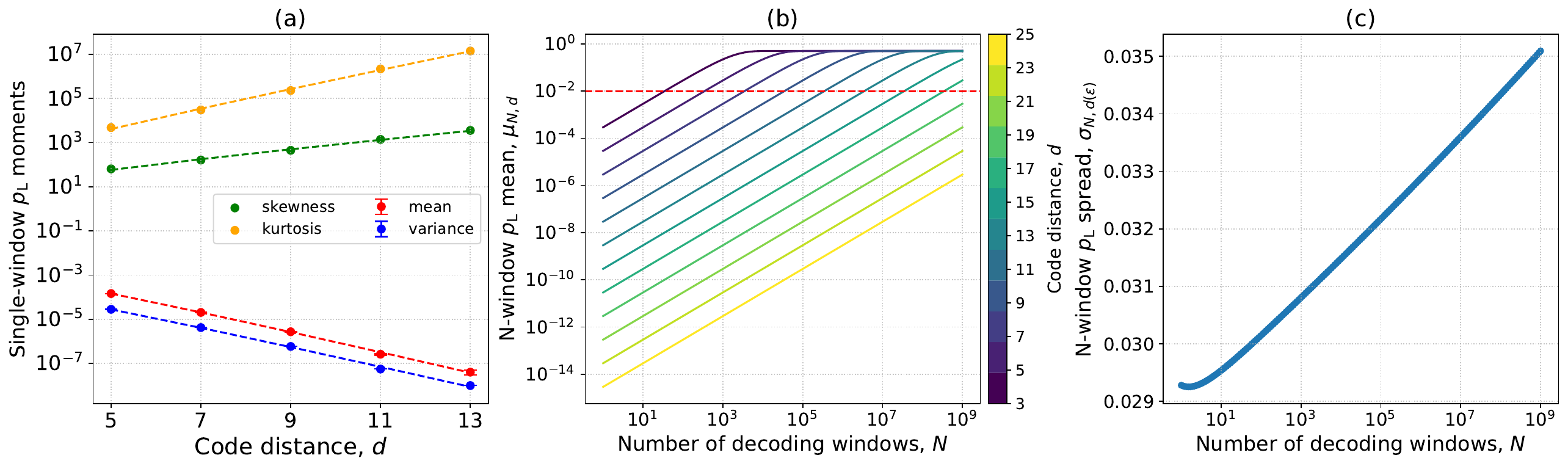}
\caption{\textbf{The moments and scaling of logical-error-probability distributions are shown across code distance and circuit size.} (a) Single-window $p_\mathrm{L}$ distribution moments as functions of code distance $d$: the legend identifies skewness, kurtosis, mean, and variance. (b) Mean whole-circuit LEP $\mu_{N,d}$ versus number of decoding windows $N$ for different code distances; the red horizontal line marks $10^{-2}$. (c) Spread $\sigma_{N,d(\epsilon)}$ when choosing the smallest code distance satisfying $\mu_{N,d(\epsilon)}\le\epsilon=0.01$; colours distinguish code distances as indicated by the colour scale.}
\label{fig:s4}
\end{figure}

\clearpage
\section*{Supplementary Results: DCS-QEM Performance Metrics}
In Figure S5,
we present an alternative performance metric for the data in Figure 7 of the main manuscript;
namely, the absolute mean bias:
\begin{equation}\label{eq:absolute_mean_bias_definition}
    \text{bias} :=
    \bigg|
    \langle Z \rangle_{\text{th}}
    -\frac{1}{N_{\text{rep}}} \sum_{i=1}^{N_{\text{rep}}}\langle Z \rangle_i
    \bigg|,
\end{equation}
which we call `Bias' for short in the y-axis labels of panels (a,c).
This metric does not account for variance,
so we plot the latter in panels (b,d).
In the limit of an infinite number of repetitions,
the noiseless estimator bias drops to exactly zero.

Panels a,c show that
MLE could achieve the target absolute mean bias
for the application described in the Hubbard-model resource estimation in the main manuscript
without any time overhead. The non-zero MLE bias is caused by finite sampling effects and imperfect LEP calibration.

Finally, in Figure S6, we showcase the results obtained for a fixed runtime budget. For each discard fraction, the retained-shot count is chosen using the time-cost factor $\omega(f)$ defined under \emph{Multiplicative Time Cost Caused by QEM} in the main manuscript, so that the mean total quantum runtime equals that of $M\in \{10^2,10^4\}$ full executions. Therefore, the final number of retained shots $M_\text{retained}$ decreases as $f$ increases. As predicted by Eq. (7) of the main manuscript, because the mean LEP $\bar{P}_\L$ also decreases with $f$, a saddle point appears in the abort protocol's MSPE evolution, which means that the increased variance introduced by fewer shots cannot always be compensated for by the reduced bias.

When the abort protocol is supplemented by MLE, this phenomenon occurs at an even lower $f$ value, because the method typically benefits from a high $P_\L$-distribution variance, which decreases as large $P_\L$ values are discarded. This raises the question: why would it ever be more advantageous to abort risky shots at a fixed runtime budget instead of maximising the information content? In general, Eq. (8) of the main manuscript does not have a closed form solution, which means that $\operatorname{Var}_{\mathrm{MLE}}$ depends on several features of the $P_\L$ distribution, not just on its first two moments. In particular, given the relatively high unmitigated LERs considered in our experiments, the $P_\L$ distribution exhibits a spike-like concentration of values close to $1/2$, as seen in Figure 8c of the main manuscript for the assumed true log odds. In Eq. (8) of the main manuscript, for shots with $P_{\mathrm{L}}\to1/2$, their contribution to the expectation value becomes much smaller than that of the other shots, which consequently increases the overall estimator variance. Therefore, it turns out that $\operatorname{Var}_{\mathrm{MLE}}$ and thus the MSPE are reduced by discarding some of these very risky shots, since they bring very little information content and time can be saved by aborting them early, if possible.

\begin{figure}[ht]
\centering
\includegraphics[width=0.98\linewidth]{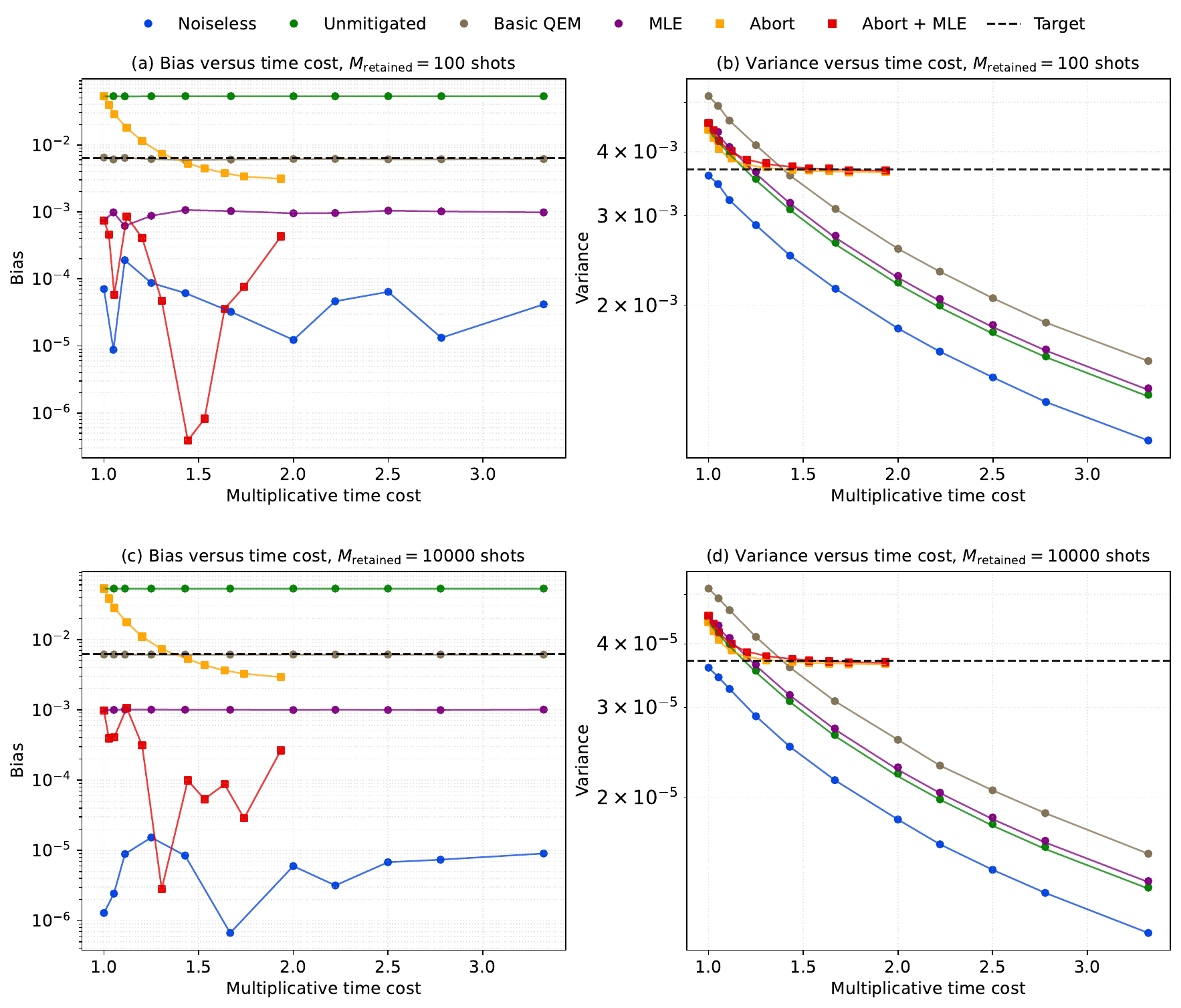}
\caption{\textbf{Bias and variance provide alternative performance metrics for the estimator data in Figure 7 of the main manuscript.} Blue denotes noiseless, green unmitigated, grey Basic QEM, purple MLE, orange Abort, red Abort+MLE, and the black dashed line the target. (a,c) Absolute mean bias versus multiplicative time cost for $M_\mathrm{retained}=100$ and $10{,}000$ shots, respectively. (b,d) Variance of the corresponding estimator output distributions.}
\label{fig:s5}
\end{figure}

\begin{figure}[ht]
\centering
\includegraphics[width=0.98\linewidth]{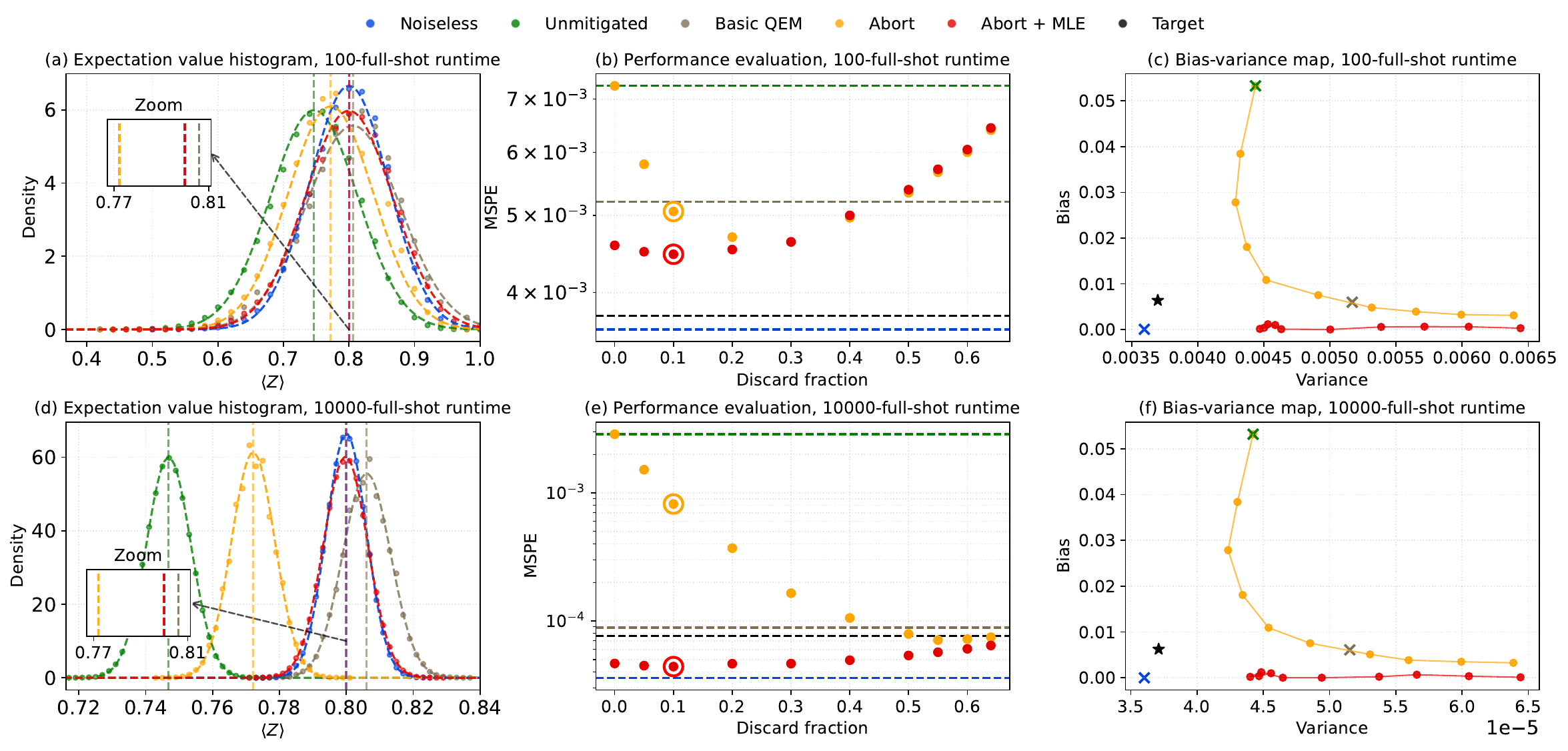}
\caption{\textbf{Estimator performance is compared at a fixed quantum-runtime budget.} Blue denotes noiseless, green unmitigated, grey Basic QEM, orange Abort, red Abort+MLE, and the black dashed line the target. (a,d) Output distributions for fixed runtime budgets equal to 100 and $10{,}000$ full executions, respectively, at a 10\% circuit discard fraction. (b,e) MSPE versus circuit discard fraction; circled points correspond to the distributions in (a,d). (c,f) Bias-versus-variance maps for the same two runtime budgets.}
\label{fig:s6}
\end{figure}


\end{document}